\documentclass[10pt,superscriptaddress,tightenlines,twocolumn,amsmath,amssymb,aps, pra]{revtex4-2}
\UseRawInputEncoding
\usepackage{graphicx}
\usepackage{dcolumn}
\usepackage{bm}
\usepackage{color}
\usepackage{amsmath}
\usepackage{amssymb}
\usepackage{subfigure}
\usepackage{braket}
\usepackage{indentfirst}
\usepackage{mathrsfs}
\usepackage[linesnumbered,ruled,vlined]{algorithm2e}
\usepackage{enumerate}
\usepackage{enumitem} 
\usepackage{algorithmic}
\usepackage{amsfonts}

\usepackage{makecell}
\usepackage{lmodern}
\usepackage{lineno}
\usepackage{multirow}
\usepackage{bbm}
\usepackage{hyperref}

\newcommand{\tn}[1]{\textnormal{#1}}
\newcommand{\be}{\begin{equation}}
\newcommand{\ee}{\end{equation}}

\newcommand{\esec}{\eps_{\tn{sec}}}

\newcommand{\sketbra}[2]{{\ensuremath{\lvert #1\rangle\!\langle #2\rvert}}}
\newcommand{\lketbra}[2]{{\ensuremath{\left\lvert #1\right\rangle\!\!\left\langle #2\right\rvert}}}
\newcommand{\ketbra}[2]{\if@display\lketbra{#1}{#2}\else\sketbra{#1}{#2}\fi}

\newcommand{\eps}{\varepsilon}

\newcommand{\cK}{\mathcal{K}}

\newcommand{\cZ}{\mathcal{Z}}

\newcommand{\rs}{{\rm{1}}}
\newcommand{\rd}{{\rm{2}}}
\newcommand{\rdd}{{\rm{3}}}

\makeatletter

\newcommand{\Rmnum}[1]{\expandafter\@slowromancap\romannumeral #1@}
\makeatother

\usepackage{xcolor}
\colorlet{RED}{red}
\colorlet{BLACK}{black}

\providecommand{\U}[1]{\protect\rule{.1in}{.1in}}

\hypersetup{colorlinks=true, linkcolor=red, citecolor=blue, urlcolor=black}

\begin{document}

\title{Scalable and highly fault-tolerant circular quantum Byzantine agreement}
\author{Chen-Xun Weng}\thanks{These authors contributed equally to this work.}
\affiliation{National Laboratory of Solid State Microstructures and School of Physics, Collaborative Innovation Center of Advanced Microstructures, Nanjing University, Nanjing 210093, China}
\affiliation{School of Physics and Key Laboratory of Quantum State Construction and Manipulation (Ministry  of  Education), Renmin University of China, Beijing 100872, China}

\author{Ming-Yang Li}\thanks{These authors contributed equally to this work.}
\affiliation{National Laboratory of Solid State Microstructures and School of Physics, Collaborative Innovation Center of Advanced Microstructures, Nanjing University, Nanjing 210093, China}
\affiliation{School of Physics and Key Laboratory of Quantum State Construction and Manipulation (Ministry  of  Education), Renmin University of China, Beijing 100872, China}

\author{Shi-Gen Li}
\affiliation{National Laboratory of Solid State Microstructures and School of Physics, Collaborative Innovation Center of Advanced Microstructures, Nanjing University, Nanjing 210093, China}
\affiliation{School of Physics and Key Laboratory of Quantum State Construction and Manipulation (Ministry  of  Education), Renmin University of China, Beijing 100872, China}

\author{Mengya Zhu}
\affiliation{MatricTime Digital Technology Co. Ltd., Nanjing 211899, China}

\author{Xiao-Ran Sun}
\affiliation{National Laboratory of Solid State Microstructures and School of Physics, Collaborative Innovation Center of Advanced Microstructures, Nanjing University, Nanjing 210093, China}
\affiliation{School of Physics and Key Laboratory of Quantum State Construction and Manipulation (Ministry  of  Education), Renmin University of China, Beijing 100872, China}

\author{Hua-Lei Yin}\email{hlyin@ruc.edu.cn}
\affiliation{School of Physics and Key Laboratory of Quantum State Construction and Manipulation (Ministry  of  Education), Renmin University of China, Beijing 100872, China}
\affiliation{National Laboratory of Solid State Microstructures and School of Physics, Collaborative Innovation Center of Advanced Microstructures, Nanjing University, Nanjing 210093, China}

\author{Zeng-Bing Chen}\email{zbchen@nju.edu.cn}
\affiliation{National Laboratory of Solid State Microstructures and School of Physics, Collaborative Innovation Center of Advanced Microstructures, Nanjing University, Nanjing 210093, China}
\affiliation{MatricTime Digital Technology Co. Ltd., Nanjing 211899, China}

\date{\today}
\begin{abstract}
Quantum Byzantine agreement (QBA), a cornerstone of quantum blockchain, offers inherent advantages in security and fault tolerance over classical protocols, guaranteed by the laws of quantum mechanics. However, existing multiparty QBA protocols face challenges for large-scale deployment due to exponential communication complexity or reliance on complex multi-particle entanglement. To address this, we propose a multiparty circular QBA protocol that adopts a semi-decentralized architecture, leveraging circular message gathering and quantum digital signatures to achieve quadratic communication complexity and enhanced fault tolerance. Our protocol is experimentally feasible, requiring only weak coherent states, and is compatible with existing star-shaped quantum networks. Simulations conducted on a global satellite-to-ground network demonstrate that the protocol sustains high consensus rates among multiple users, even when employing different key generation protocols under realistic conditions. This work presents a scalable framework for large-scale QBA networks, establishing the foundation for a practical quantum blockchain that enables secure and fault-tolerant decentralized services.
\end{abstract}

\maketitle

\section{Introduction}
Originating from the Byzantine Generals Problem~\cite{lamport1982byzantine}, Byzantine agreement protocols enable all honest nodes to reach consensus on a value even in the presence of malicious nodes~\cite{castro1999practical,aublin2013RBFT,guo2020dumbo}. This task is fundamental to decentralized applications such as blockchain, distributed database and the metaverse~\cite{bodkhe2020Blockchain,jiang2024blockchain,miao2026blockchain}. Classical protocols rely on computational assumptions that are increasingly threatened by advances in quantum algorithms and quantum computers
~\cite{shor1999polynomial,grover1997quantum,fedorov2018quantum,zhou2022experimental}. In addition, classical protocols where nodes are linked via point-to-point channels are rigorously constrained by the 1/3 fault-tolerance bound, which means the system could tolerate at most 1/3 nodes is malicious~\cite{dolev1986possibility,fischer1986easy,feng2026pbft}.

Quantum Byzantine agreement (QBA) leverages quantum resources to provide quantum advantages in both security and fault tolerance for consensus mechanisms. The first QBA protocol, based on entangled qutrit states (Aharonov state), is proposed to achieve a three-party consensus that is unattainable by classical protocols~\cite{Fitzi2001quantum}, and is experimentally demonstrated using tailored four-photon polarization-entangled states~\cite{gaertner2008xperimental}. Currently, feasible detectable QBA protocols remain limited to three-party consensus and lack a clear, practical framework for extending beyond three-party scenarios because of their reliance on multi-particle entanglement or high-dimensional qudits~\cite{Iblisdir2004byzantine,Neigovzen2008Multipartite,Rahaman2015Quantum,smania2016experimental}. Generally, a ``detectable'' QBA protocol satisfies Byzantine conditions only when it terminates successfully, as honest players will abort the process upon detecting malicious inconsistencies. Recently, a multiparty recursive QBA protocol has been proposed~\cite{weng2023beating}, which exploits recursive structures together with one-time universal hashing quantum digital signatures (OTUH-QDS)~\cite{yin2023experimental,li2023one,xiong2024efficient,bian2024Asynchronous,qin2024efficient,du2025chip}. The term ``recursive'' refers to a layered message-passing structure where a three-party OTUH-QDS is recursively utilized to relay signed messages layer by layer throughout the network to ensure consistency.  Unlike detectable approaches relying on multipartite entangled states or high-dimensional qudits, recursive QBA requires only weak coherent states and can, in principle, be extended to an arbitrary number of participants. Owing to the asymmetric properties of three-party QDS~\cite{gottesman2001quantum,dunjko2014quantum,yin2016practical,amiri2016secure,roberts2017experimental,richter2021agile,qin2022quantum}, the protocol achieves information-theoretic security and a fault tolerance approaching one half, thereby surpassing the classical 1/3 fault-tolerance bound and demonstrating a quantum advantage in distributed consensus. It has been experimentally demonstrated in a three-party setting using an integrated photonic circuit~\cite{jing2024experimental}, and further extended to a five-party scenario based on a fully heterogeneous prepare-and-measure quantum network~\cite{Lu2025Fully}.

With the rapid development of quantum networks~\cite{Wehner2018Quantum,Azuma2023Quantum,bozzio2025quantum}, it is intriguing to explore whether multiparty QBA can be deployed on large-scale quantum networks for practical applications, such as metropolitan and satellite-to-ground quantum networks. If realized, such a QBA network could serve as a core consensus mechanism for constructing quantum blockchains that provide information-theoretically (i.-t.) secure and highly fault-tolerant decentralized information services for real-world applications, such as quantum-blockchain-based federated learning~\cite{liu2025quantum}, distributed data trust management~\cite{qu2023quantum}, distributed finance~\cite{Liu2025Experimental}, internet of things~\cite{zhao2023secure}, and secure authenticated message delivery in vehicular networks~\cite{Dharminder2021edge,Kumar2025NTRU,Singh2025Building}. 

However, current QBA protocols face three main limitations for large-scale deployment. (1) Detectable QBA requires multi-particle entanglement or high-dimensional qudits, making it difficult to scale beyond three parties~\cite{sun2020multi,li2024quantum,Paing2024Counterfactual}. (2) Although recursive QBA~\cite{weng2023beating} and QKD-based QBA~\cite{kiktenko2018quantum} use practical weak coherent states, their communication complexity scales exponentially with the number of participants. This occurs because both methods rely on recursive structures, where the recursion depth depends on the number of malicious nodes. (3) Current protocols all  assume a fully connected network. For an $N$-party system using point-to-point QKD, this requires $N(N-1)/2$ quantum channels. In practice, large-scale quantum networks usually use star or ring topologies rather than full connections.

To address these issues, we propose a semi-decentralized, i.-t. secure circular QBA protocol. Our motivation is to provide a consensus mechanism that is secure, scalable, and compatible with existing quantum network architectures. Our main contributions are as follows. (1) Our protocol uses OTUH-QDS based on weak coherent states instead of multi-particle entanglement, which enables efficient multi-bit signature generation and verification. It requires only two honest nodes to reach consensus (i.e., $N \ge f + 2$, where $f$ is the number of malicious nodes). (2) We replace the recursive structure with a circular message-gathering process. This reduces the communication complexity from exponential to polynomial $\mathcal{O}(N^2)$ while ensuring unforgeability and non-repudiation. (3) Our semi-decentralized architecture simplifies the network topology by requiring only $N$ quantum channels to distribute keys for OTUH-QDS. This design fits satellite-to-ground quantum networks, where the satellite acts as a certificate authority (CA). Simulations show that our protocol achieves a high consensus rate under realistic satellite-to-ground conditions, demonstrating its feasibility for practical deployment.

\begin{figure*}[t]
    \centering
    \includegraphics[width=\linewidth]{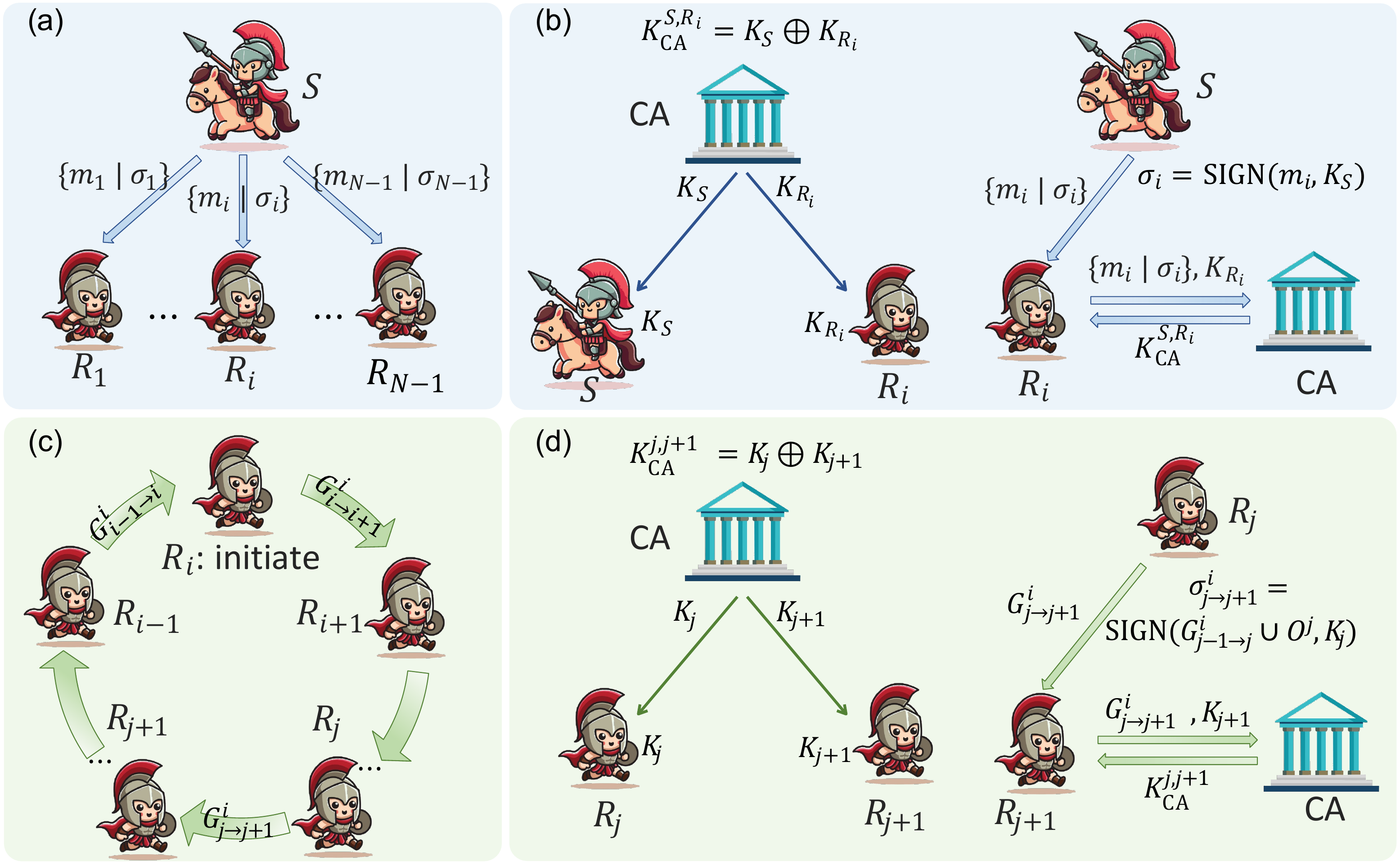}
    \caption{Schematic of our circular QBA protocol. (a) Order distribution. The commanding general $S$ distributes message $m_i$ to each lieutenant $R_i$ with signature $\sigma_i$ through three-party OTUH-QDS ($\forall i \in \mathbb{Z}_{N-1}^+$). (b) Message distribution step between $S$, $R_i$, and CA in the order distribution phase. $S$, $R_i$, and the CA share correlated quantum keys through a KGP first (left). Then they conduct the signing and verifying procedure (right).  $R_i$'s initial message list is $O^i=\{m_i,\sigma_i\}$. CA records the signature $\sigma_i$ and related quantum keys of this QDS. (c) Circular gathering started by $R_i$ ($\forall i \in \mathbb{Z}_{N-1}^+ $). The gathering follows a clockwise sequence, and each step is completed using OTUH-QDS. (d) The information delivery step from $R_{j}$ to $R_{j+1}$. $R_{j}$ combines the information package received from the previous lieutenant $R_{j-1}$, which is $G^i_{j-1}$, and its initial message list, which is $O^j$, and takes it as a whole and signs it with a signature $\sigma^i_{j\rightarrow j+1}$. Thus $G^i_{j\rightarrow j+1} = G^i_{j-1\rightarrow j} \cup   O^j \cup \{\sigma^i_{j\rightarrow j+1}\}$. The OTUH-QDS process is similar to those in order distribution. CA acts as the verifier of QDS, records $\sigma^i_{j\rightarrow j+1}$ and checks the validity of all signatures in $G^i_{j\rightarrow j+1}$. Icons Copyright 2026 Vecteezy. 
}
    \label{Fig1}
\end{figure*}

\section{Semi-decentralized QBA design}
Our QBA protocol employs the three-party OTUH-QDS~\cite{yin2023experimental,li2023one} as the core component, which provides i.-t. secure non-repudiation and unforgeability. OTUH-QDS involves three roles: signer ($\mathbb{S}$), forwarder ($\mathbb{F}$), and verifier ($\mathbb{V}$), with two phases: the distribution phase and the messaging phase. In the distribution phase, the three parties utilize a key generation protocol (KGP, see Appendix~\ref{KGP}) such as QKD or quantum secret sharing to establish correlated secret key strings, denoted as $\{ X_{k}, Y_{k}, Z_{k} \}$ for user $k \in \{\text{$\mathbb{S}$, $\mathbb{F}$, $\mathbb{V}$}\}$. These keys satisfy the XOR correlation where the signer's key equals the bitwise sum of the others: $X_\mathbb{S}=X_\mathbb{F}\oplus X_\mathbb{V}$, $Y_\mathbb{S}=Y_\mathbb{F}\oplus Y_\mathbb{V}$, and $Z_\mathbb{S}=Z_\mathbb{F}\oplus Z_\mathbb{V}$. 

In the messaging phase, the signer defines a Linear-feedback shift register (LFSR)-based Toeplitz hash function $H_{X_{\mathbb{S}},p_{\rm{s}}}$ determined by key $X_{\mathbb{S}}$ and a random parameter $p_{\rm{s}}$. For a message $\mathrm{Mes}$, the signer computes the digest $\mathrm{Dig}=H_{X_{\mathbb{S}},p_{\rm{s}}}(\mathrm{Mes})$. This digest is encrypted to generate the signature $\mathrm{Sig}=\mathrm{Dig}\oplus Y_{\mathbb{S}}$, while $p_{\rm{s}}$ is encrypted as $p=p_{\rm{s}}\oplus Z_{\mathbb{S}}$. The signer transmits $\{\mathrm{Mes}, \mathrm{Sig}, p \}$ to the forwarder.

Upon receiving the data, the forwarder transmits $\{\mathrm{Mes}, \mathrm{Sig}, p \}$ along with keys $\{ X_{\mathbb{F}}, Y_{\mathbb{F}}, Z_{\mathbb{F}} \}$ to the verifier. The verifier recovers the signer's keys by calculating the XOR between the received keys and their own (e.g., $K_{X_{\mathbb{V}}} = X_{\mathbb{V}} \oplus X_{\mathbb{F}}$) and verifies the signature by comparing the decrypted digest against the locally computed hash of $\mathrm{Mes}$. If successful, the verifier sends keys $\{ X_{\mathbb{V}}, Y_{\mathbb{V}}, Z_{\mathbb{V}} \}$ to the forwarder. The forwarder then reconstructs the signer's keys (e.g., $K_{Y_{\mathbb{F}}} = Y_{\mathbb{F}} \oplus Y_{\mathbb{V}}$), decrypts $\mathrm{Sig}$ and $p$, and computes the actual digest using the hash function derived from $K_{X_{\mathbb{F}}}$. The signature is accepted only if both the forwarder and the verifier confirm that the computed digest matches the decrypted one.

In our scalable multi-party QBA framework, we introduce a designated entity, the certificate authority (CA), which exclusively assumes the role of the verifier in every QDS transaction. The role of the CA is strictly confined to providing a cryptographic service: it verifies the validity of signatures upon request from the players. Crucially, the CA serves as an impartial assistant that does not actively participate in the consensus process. It neither generates nor relays messages between players, contributes to the final decision, nor provides any information to a player other than the binary result of a signature verification (i.e., valid or invalid). Such an assumption is also consistent with classical Byzantine agreement protocols that rely on digital signatures, where a trusted setup such as a public-key infrastructure is typically required to distribute and authenticate public keys~\cite{lamport1982byzantine,castro1999practical}. In this sense, the CA in our protocol plays a similar role as a supporting cryptographic infrastructure, rather than a decision-making entity. This architecture leads to our \textit{semi-decentralized} model, where we achieve a profound simplification in network topology and a reduction in communication complexity, thereby enabling scalability that would be difficult in a fully decentralized QBA network. The detailed steps and security analysis of OTUH-QDS can be found in Appendix~\ref{OTUH}.

\section{Protocol description}
The Byzantine Generals problem is typically framed with a commanding general $S$ and several lieutenants $R_i$~\cite{lamport1982byzantine}. A protocol is considered successful only if it satisfies two Interactive Consistency (\textbf{IC}) conditions: (1) All honest lieutenants agree on the same value (\textbf{IC$_1$}); and (2) If the commanding general is honest, then all honest lieutenants adopt the correct general's value (\textbf{IC$_2$}).

In our circular QBA, a commanding general $S$ is randomly selected from $N$ participants, with the other $N-1$ serving as lieutenants $R_i$  ($i \in \mathbb{Z}_{N-1}^+$, where $\mathbb{Z}_{N-1}^+ = \{1,2,\cdots,N-1\}$). The protocol has three phases: order distribution, where the general sends orders to lieutenants; circular gathering, where lieutenants collect messages from each other; and consensus output, where each honest lieutenant independently outputs the final value according to what they collected before. Figure~\ref{Fig1} illustrates the steps of our circular QBA. The details of the three phases are as follows.

\textbf{\textit{1. Order distribution.} } For each lieutenant $R_i$ ($i \in \mathbb{Z}_{N-1}^+$), the commanding general $S$ signs an $m$-bit order $m_i$ with an $n$-bit signature $\sigma_i$ in Fig.~\ref{Fig1}(a). This pair $\{m_i;\sigma_i\}$ is sent to $R_i$ via a three-party OTUH-QDS, with $S$, $R_i$, and the CA as the signer, forwarder, and verifier. Each OTUH-QDS session is shown in Fig.~\ref{Fig1}(b): first, a KGP where $S$, $R_i$, and the CA establish their correlated keys; second, the classical signing and verification procedure. Upon verification completes, $R_i$ stores its order list $O^{i} = \{m_i;\sigma_i\}$.  If the general is loyal, all orders received by $R_i$ are identical: $m_i=m_0~(\forall i\in \mathbb{Z}_{N-1}^+)$.

\textbf{\textit{2. Circular gathering.}} Each of the $N-1$ lieutenants initiates a clockwise circular gathering of $N-1$ steps (Fig.~\ref{Fig1}(c)). We take the circular gathering started by $R_i$ as an example. The gathering follows a clockwise cycle, denoted by the sequence $(i \rightarrow i+1\rightarrow \cdots\rightarrow N-1 \rightarrow 1 \rightarrow \cdots \rightarrow i)$. The first information package delivered from $R_i$ to $R_{i+1}$ is $R_i$'s initial message list $O^i$ signed with $\sigma^i_{i\rightarrow i+1}$, which is denoted as $G^i_{i\rightarrow i+1}=\{m_i;\sigma_i;\sigma^i_{i\rightarrow i+1}\}$. In the clockwise cycle, one of the participants, lieutenant $R_j$ $(j\neq i)$, who receives $G^i_{j-1\rightarrow j}$ from $R_{j-1}$ and sends $G^i_{j\rightarrow j+1}$ to $R_{j+1}$, combines $G^i_{j-1\rightarrow j}$ with its own initial message list ($O^{j}=\{m_j;\sigma_j\}$), and signs them with signature $\sigma^i_{j\rightarrow j+1}$, which means that $G^i_{j\rightarrow j+1}= G^i_{j-1\rightarrow j}  \cup O^j \cup \{\sigma^i_{j\rightarrow j+1}\} = \{m_i,\cdots,m_j;\sigma_i,\cdots,\sigma_j;\sigma^i_{i\rightarrow i+1},\cdots,\sigma^i_{j\rightarrow j+1} \}$. The information packages are delivered via three-party OTUH-QDS as shown in Fig.~\ref{Fig1}(d), which is similar to the QDS in the order distribution. 

After the clockwise gathering, $R_i$ will receive the final list $G^i_{i-1\to i}$ which contains three parts: (1) all the lieutenants' received order $\{m_1,m_2,\dots,m_{N-1}\}$. (2) the signatures signed by the commanding general $\{\sigma_1,\sigma_{2},\dots,\sigma_{N-1}\}$, and (3) the signatures generated during the circular gathering by lieutenants $\{\sigma^i_{i\to i+1},\sigma^i_{i+1\to i+2},\dots,\sigma^i_{N-1\to 1},\dots,\\ \sigma^i_{i-1\to i}\}$.

Note that in each step of circular gathering, CA verifies all the signatures in the information package. First, CA verifies the validity of $\sigma^i_{j\rightarrow j+1}$ as the regular QDS process and records it if valid. Second, CA further verifies that the signatures of all previous steps are unchanged and valid according to the record of CA, i.e., $\{\sigma_i,\cdots,\sigma_j;\sigma^i_{i\rightarrow i+1},\cdots,\sigma^i_{j-1\rightarrow j}\}$. Because CA maintains the records of correlated quantum keys and signatures of each OTUH-QDS process, the verification is straightforward. Any failure of verification will restart a new round of circular gathering initiated by $R_i$. 

\textbf{\textit{3. Consensus output.}} After the circular gathering, each lieutenant $R_i$ $(\forall i \in \mathbb{Z}_{N-1}^+)$ gets the completely identical message list, namely $F^i = \{ m_1, m_2, \dots, m_{N-1} \}$, and puts it into a pre-selected deterministic function $D$ to produce the final output $m^{R_i}=D(m_1, m_2, \dots, m_{N-1})$. Note that a deterministic function is a function that always produces the same output for the same input.

\section{Security analysis of circular QBA}

In this part, we analyze the i.-t. security and fault-tolerance ability of our circular QBA protocol. The details of the security analysis can be found in Appendix~\ref{securityAnalysis}.

\textbf{(a) \textit{Overall security.}} The security of our protocol rests on the i.-t. secure non-repudiation and unforgeability of its underlying cryptographic primitive, the OTUH-QDS. As demonstrated in Ref.~\cite{yin2023experimental}, repudiation is impossible, while the probability of a successful forgery is bounded by
\begin{equation}
        \varepsilon_{\rm{for}}(\mathcal{M}, \mathcal{N}) =\mathcal{M}  \cdot2^{1-\mathcal{N}},
    \label{parameter}
\end{equation}
where $\mathcal{M}$ and $\mathcal{N}$ are the lengths of the message and signature, respectively.
 
To ensure the protocol satisfies the two \textbf{IC} conditions, all QDS steps must be secure against forgery. The overall failure probability of the QBA, $\varepsilon_{\rm{QBA}}$, is therefore determined by the cumulative risk of forgery across all vulnerable communication steps, i.e., all the QDS steps with dishonest nodes. Crucially, the forgery probability in Eq.~(\ref{parameter}) depends on the message length $\mathcal{M}$, which varies and grows during the circular gathering phase. Therefore, our security analysis adopts a conservative, worst-case approach: we assume malicious players will always target the QDS steps involving the longest possible messages. The final bound also accounts for different scenarios based on the general's honesty, as this affects the number of malicious lieutenants. This reasoning leads to an upper bound on the total failure probability, expressed as
\begin{equation}
    \begin{aligned}
        \varepsilon_{\rm{QBA}}        =&\max \{ f\left [  \varepsilon_{\rm{for}}(m, n) + \left (N-f-1\right ) \cdot \varepsilon_{\rm{for}}(L_{N-1},n) \right ],\\
        &(f-1)(N-f)\cdot \varepsilon_{\rm{for}}(L_{N-1},n)\},
    \end{aligned}
\end{equation}
where $L_{N-1} = (N-1)m + (2N-3)n$ is the length of the longest message signed in a circular gathering. A detailed derivation of this bound is provided in Appendix~\ref{securityAnalysis}.

Furthermore, we consider composable security by accounting for imperfections in the KGPs used to establish the shared keys~\cite{yin2017experimental,korzh2015provably}. The overall security of the protocol is thus modeled by the sum of the protocol's QBA failure probability and that of the underlying KGPs,
\begin{equation}
\varepsilon_{\text{all}}=\varepsilon_{\text{QBA}} + \varepsilon_{\text{KGP}},
\end{equation}
where $\varepsilon_{\text{KGP}}$ represents the total failure probability of all KGPs involved.

\textbf{(b) \textit{Fault-tolerance.}} Here we consider two cases where the commanding general $S$ is honest or not according to $\text{\textbf{IC}}_1$ and $\text{\textbf{IC}}_2$. When $S$ is honest, the distribution of the message $m_0$ is guaranteed with unforgeability because at least two honest participants ($S$ and CA) recording the message distribution and its signature. Only the correct message $m_0$ appears in the circular gathering phase and ensures that all honest lieutenants reach a unanimous decision based on $N-1$ copies of $m_0$. When $S$ is dishonest, although $N-1$ distinct messages $\{ m_1,m_2,\cdots,m_{N-1} \}$ are distributed, the CA still records them to prevent any forgeries when an honest lieutenant performs a circular gathering. All honest lieutenants gather the same set of messages $\{ m_1,m_2,\cdots,m_{N-1} \}$. This ensures all honest lieutenants output the same result. As a result, our protocol requires at least two honest participants in the system, which means that~\nocite{liao2017satellite,Chen2021AnIntegrated,li2025microsatellite}
\begin{equation}
    N \geq f + 2.
\end{equation}

\begin{figure}[t]
    \centering
    \includegraphics[width=\linewidth]{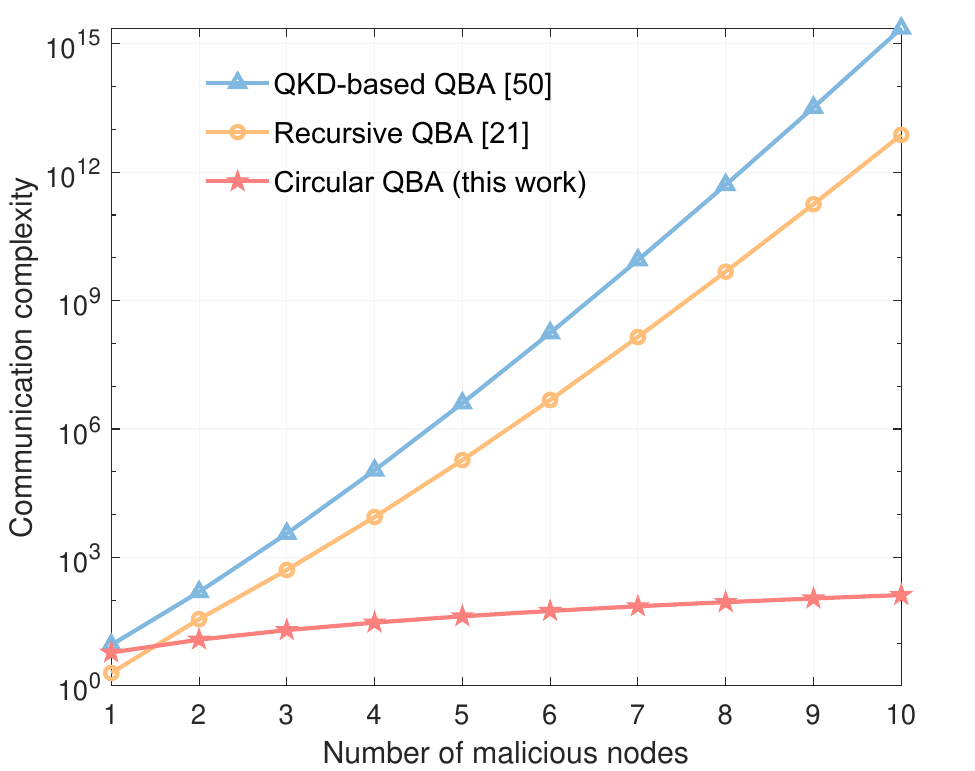}
    \caption{Comparison of lower bounds on communication complexity (number of communication rounds) for QKD-based QBA~\cite{kiktenko2018quantum}, recursive QBA~\cite{weng2023beating} and this work. For each protocol, the total number of nodes is minimized according to its fault-tolerance capability. For the number of malicious players $f \geq 2$, our protocol provides an obvious advantage over the previous schemes.
}
    \label{Fig2}
\end{figure}

\begin{figure}[t]
    \centering
    \includegraphics[width=\linewidth]{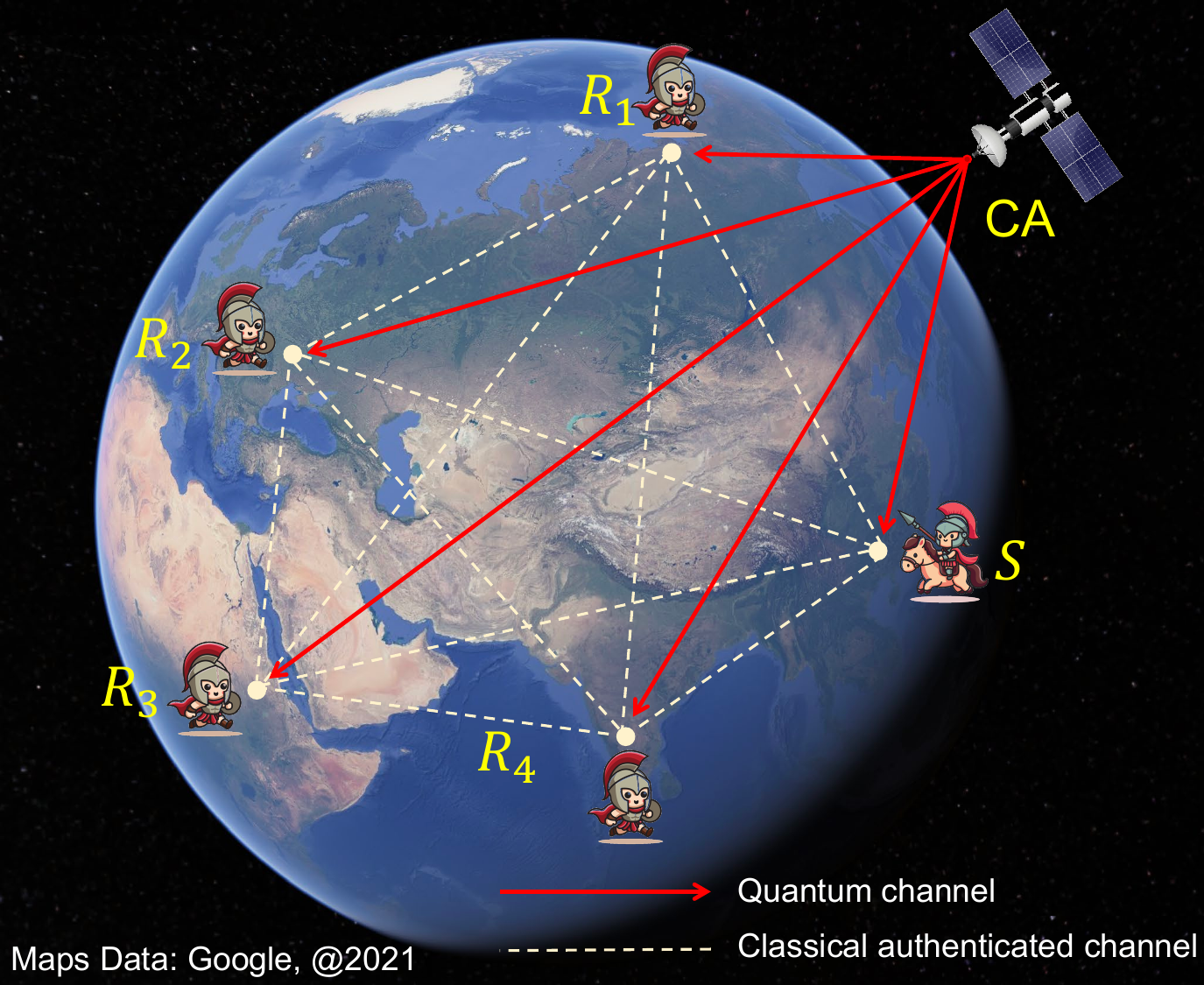}
    \caption{Illustration of the satellite-to-ground five-user QBA network employing our circular QBA protocol. The satellite acts as the CA, distributing quantum keys to ground-based players via satellite-to-ground quantum channels (red solid lines) and verifying validity in each three-party QDS. Authenticated classical channels (yellow dashed lines) are employed for communication among users, avoiding the need for quantum channels between ground stations. Note that classical authenticated channels between a player and CA are needed for the communication during QDS. Map data Copyright 2021 Google. Icons Copyright 2026 Vecteezy.}
    \label{Fig3}
\end{figure}

\begin{figure*}[t]
    \centering
    \includegraphics[width=\linewidth]{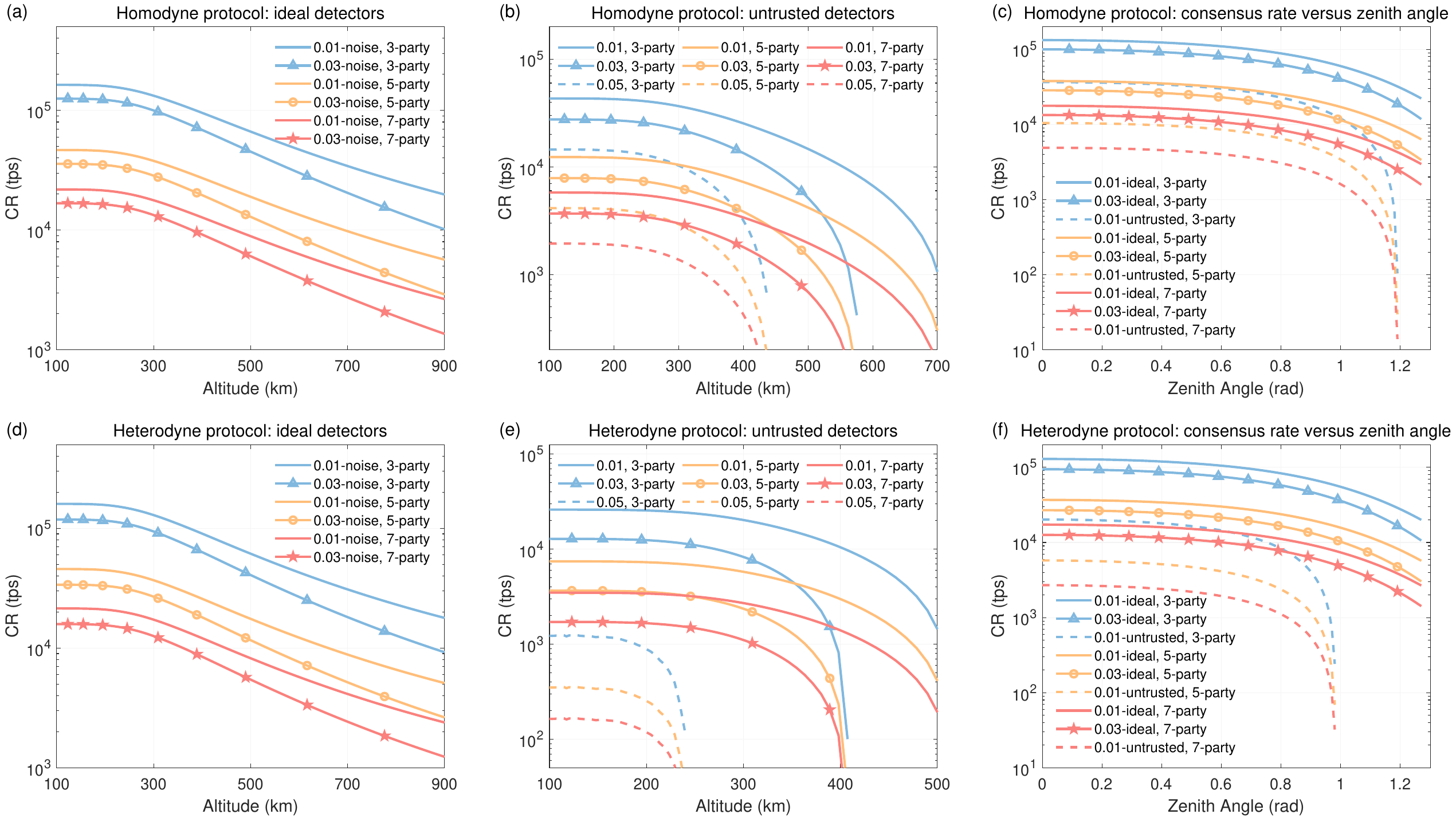}
    \caption{Simulation of the consensus rate of circular QBA using satellite-to-ground DM-CV KGP scheme~\cite{Li2024Discrete}. Figures (a), (b), and (c) present the consensus rate versus altitude of the satellite (CA) at a zenith angle of 0 rad with ideal detectors, untrusted detectors, and the consensus rate versus the zenith angle at an altitude of 300 km, when using the homodyne protocol~\cite{liu2021homodyne}. Figures (d), (e), and (f) are the corresponding results when using the heterodyne protocol~\cite{lin2019Asymptotic}. All simulations are performed under downlink conditions, as the proposed QBA scheme requires the CA to distribute quantum keys. We consider different noise levels ($\xi_{\rm{ch}} = 0.01,~0.03,~0.05$) with both ideal and untrusted detectors. The size of the message is fixed to 1 Mb. These results comprehensively demonstrate the high efficiency of our protocol in a satellite-to-ground system.
}
    \label{Fig4}
\end{figure*}

\begin{figure*}[t]
    \centering
    \includegraphics[width=0.9\linewidth]{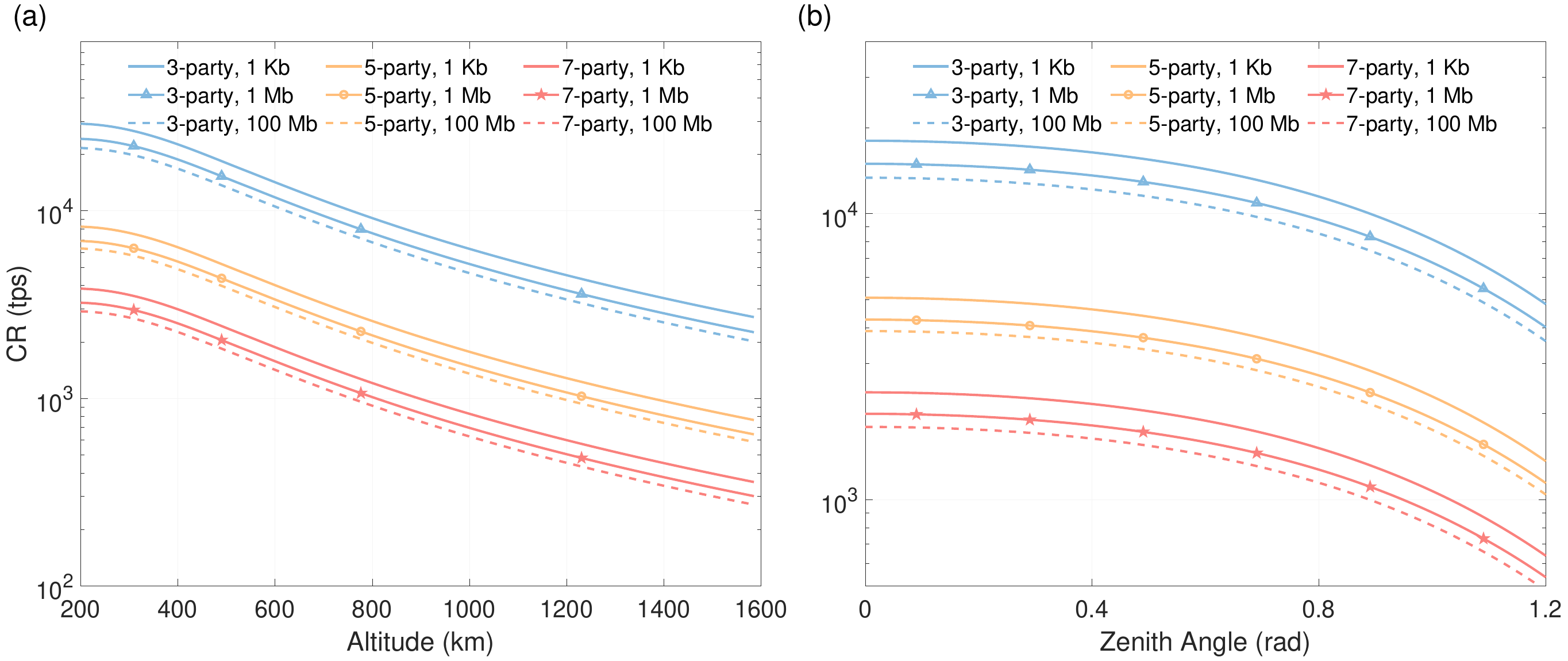}
    \caption{Performance of the QBA protocol using BB84 KGP adapted from asymmetric coding BB84 QKD~\cite{lo2005Efficient,lim2014concise}. (a) Consensus rate versus satellite altitude (zenith angle = 0 rad). (b) Consensus rate versus zenith angle (at a fixed altitude of 300 km). Results are shown for various numbers of participants and message sizes (1 Kb, 1 Mb, and 100 Mb). }
    \label{Fig5}
\end{figure*}

\section{Scalability analysis of circular QBA}

Here, we analyze the communication complexity and the Number of required quantum channels of our circular QBA protocol.

\textbf{(a) \textit{Communication complexity.}} We define the communication complexity, $C(N)$, as the minimum times of communication (QDS in our protocol) to reach consensus for a distributed system of $N$ players with $f$ malicious ones. In our work, the order distribution phase comprises $N-1$ distribution steps from commanding general $S$, and the circular gathering phase comprises $N-1$ lieutenants independently initiating circular gathering, and each circular gathering contains $N-1$ message delivery steps. Therefore, the communication complexity of circular QBA is 
\begin{equation}
    C(N)=(N-1)+(N-1)(N-1)=N^2-N.
\end{equation}
In Fig.~\ref{Fig2}, we compare the communication complexity (measured by the number of communication rounds) of our protocol with detectable QBA~\cite{Fitzi2001quantum} and two existing i.-t. secure multiparty QBA protocols, QKD-based QBA~\cite{kiktenko2018quantum} and recursive QBA~\cite{weng2023beating}. The figure presents lower bounds on communication complexity, determined by minimizing the total number of nodes while maintaining security for a given number of malicious participants. The results demonstrate that, except in the single-malicious-node case ($f=1$), our protocol exhibits significantly superior scalability and fault-tolerance, particularly for large-scale systems. Specifically, for $f=10$ malicious nodes, QKD-based QBA~\cite{kiktenko2018quantum} requires at least $2.30\times 10^{15}$ rounds, and recursive QBA protocol~\cite{weng2023beating} requires at least $7.44\times 10^{12}$ rounds, whereas our protocol requires only 132 rounds, achieving a reduction of more than ten orders of magnitude.

\textbf{(b) \textit{Number of quantum channels.}} Our protocol's architecture reduces quantum channel requirements. The three-party correlated keys needed for OTUH-QDS are established efficiently: with the CA acting as a verifier, it performs two separate QKD processes with the other participants (signer and forwarder) and locally computes its key via bitwise XOR. This star-shaped model simplifies the network from a fully-connected mesh requiring $N(N-1)/2$ quantum channels~\cite{kiktenko2018quantum,weng2023beating} to a star-shaped network with only $N$ channels connecting players to the CA. Such a configuration is well-suited for a satellite-based implementation, where the satellite naturally serves as the central CA, overcoming the attenuation limits of terrestrial fiber. We present this satellite architecture in detail in the following section.

\section{Satellite-based large-scale circular QBA networks}

The successful demonstration of satellite-to-ground QKD has established a viable path towards a global quantum internet, overcoming the distance limitations imposed by fiber attenuation~\cite{liao2017satellite,Chen2021AnIntegrated,li2025microsatellite}. However, these pioneering networks fundamentally serve for some information tasks involving trusted nodes, primarily for point-to-point secure links. While sufficient for quantum key distribution involving two trusted parties, this poses a critical limitation on multiparty distributed applications, such as blockchain where nodes may be untrusted with each other. The next step for the applications of quantum networks, whether terrestrial or space-based, is to develop towards fault-tolerant decentralized systems. QBA is the foundational protocol for achieving such system. With a polynomial communication complexity and linear scaling of the number of quatnum channels, our circular QBA protocol is practical when being applied in large-scale satellite-to-ground star-shaped quantum networks.

Satellite-to-ground communication involves signal transmission through the atmosphere instead of optical fibers, enabling long-distance communication~\cite{Li2024Discrete}. Depending on the receiver's position, the channel is categorized as either downlink or uplink. Here we focus on the downlink, where the receiver (players) is located on the ground, as shown in Fig.~\ref{Fig3}. Atmospheric effects on the optical signal can be classified into attenuation, turbulence, and deflection. Attenuation is primarily caused by aerosol and molecular scattering, while turbulence and deflection result from the uneven spatiotemporal distribution of the atmosphere. As the laser beam travels through the atmosphere, this unevenness causes the beam waist to broaden and the beam's direction to deviate randomly. All of these effects, together with possible laser misalignment, contribute to a reduction in total transmittance, which is described in Appendix~\ref{Simulation details}. The extent of these effects depends on many factors, most of which are determined by the environment. Here, we focus on the satellite's altitude and zenith angle in subsequent simulations, considering controllability.

Here, to present the efficiency, we simulate the consensus rate ($CR$) of our protocol, defined as the number of successful consensus that can be achieved per second. Since OTUH-QDS are required for each consensus at least $C(N) = N^2 - N$ times, the $CR$ is proportional to the signature rate ($SR$, see Appendix~\ref{OTUH} for details). $CR$ can be expressed as~\cite{weng2023beating}
\begin{equation}
    CR = \frac{\min(SR)}{C(N)} = \frac{\min(SR)}{(N^2 - N)},
\end{equation}
where the minimum function ($\min$) accounts for variations in the quantum channels between parties, leading to different channel losses and varying signature rates.

\begin{table*}[t]
\renewcommand\arraystretch{1.6}
\caption{\label{Tabel2}Comparison between our protocol and other QBA protocols. We mainly consider three aspects: security, scalability, and decentralization. Detectable: detectable QBA has a nonzero probability of avoidable failure, where the events should be disregarded. $P^n_r = \frac{n!}{(n - r)!}$ is $r$-permutations of $n$. N/A: not applicable.}
\resizebox{\linewidth}{!}{ 
\begin{tabular}{llcccccc}
\hline\hline

\multicolumn{2}{c}{\textbf{Performance index}} & \makecell{\bf Fitzi \emph{et~al.}\\\cite{Fitzi2001quantum}} & \makecell{\bf Rahaman \emph{et~al.}\\\cite{Rahaman2015Quantum}} & \makecell{\bf Smania \emph{et~al.}\\\cite{smania2016experimental}} & \makecell{\bf Kiktenko \emph{et~al.}\\\cite{kiktenko2018quantum}} & \makecell{\bf Weng \emph{et~al.}\\\cite{weng2023beating}} & \makecell{\bf This work} \\ 
\hline

\multirow{2}*{Security}&\makecell{Fault-tolerance} & \makecell{$f=1,N=3$} & \makecell{$f=1,N=3$} & \makecell{$f=1,N=3$} & \makecell{$N\ge 3f+1$} & \makecell{$N\ge 2f+1$} & \makecell{$N\ge f+2$} \\ 
\cline{2-8}

~&\makecell{Security level} & \makecell{Detectable} & \makecell{Detectable} & \makecell{Detectable} & \makecell{i.-t. secure} & \makecell{i.-t. secure} & \makecell{i.-t. secure} \\ 
\hline

\multirow{4}{*}{Scalability}&\makecell{Communication \\ complexity} & \makecell{N/A} & \makecell{N/A} & \makecell{N/A} & \makecell{$\sum_{r=1}^{f+1} P^{N-1}_{r}$} & \makecell{$\sum_{r=0}^{f-1} P^{N-1}_{r+2}$} & \makecell{$N^2-N$} \\ 
\cline{2-8}

~&\makecell{Number of \\ quantum channels} & \makecell{3} & \makecell{3} & \makecell{3} & \makecell{$N(N-1)/2$} & \makecell{$N(N-1)/2$} & \makecell{$N$} \\
\cline{2-8}

~&\makecell{Experimental \\ requirement} & \makecell{Entangled \\ Aharonov states} & \makecell{Entangled \\ Hardy states} & \makecell{Single \\ qutrit} & \makecell{Coherent \\ states} & \makecell{Coherent \\ states} & \makecell{Coherent \\ states} \\
\cline{2-8}

~&\makecell{Message type} & \makecell{Single bit} & \makecell{Single bit} & \makecell{Single bit} & \makecell{Single bit} & \makecell{Multi-bit} & \makecell{Multi-bit} \\
\hline

\multicolumn{2}{l}{Decentralization} & \makecell{Full} & \makecell{Full} & \makecell{Full} & \makecell{Full} & \makecell{Full} & \makecell{Semi} \\ 
\hline

\multicolumn{2}{l}{Cryptography tool} & \makecell{Quantum \\ entanglement} & \makecell{Quantum \\ entanglement} & \makecell{High-dimension \\ quantum state} & \makecell{QKD} & \makecell{QDS} & \makecell{QDS} \\ 
\hline
\hline
\end{tabular}}
\end{table*}

We first simulate the performance of our circular QBA using the satellite-to-ground discrete-modulated continuous-variable (DM-CV) QKD~\cite{Li2024Discrete}, under the conditions of both homodyne detection~\cite{liu2021homodyne} and heterodyne detection~\cite{lin2019Asymptotic}. In these simulations, we optimize the performance by setting the intensity of the coherent state $\alpha = 0.72$ and the post-selection parameters $\Delta_c=0.42 $, $\Delta_a=0.52$, and $\Delta_p=0$. The repetition rate is $10^9$ Hz, and the detectors' noise is $\xi_{\rm{het}} = 2\xi_{\rm{hom}} = 2\xi_{\rm{det}} = 0.02$. Figure~\ref{Fig4} presents the consensus rate achieved using the two detection methods with different numbers of participants, noise levels, and the types of detectors (trusted or untrusted).  All simulations are conducted under the downlink condition to model the scenario where the satellite (CA) distributes the quantum keys. The results demonstrate high efficiency under various conditions, especially for Low Earth Orbit (LEO) satellites with an altitude of around 200-900 km. A consensus rate of more than $10^2$ to $10^5$ times per second can be achieved for multi-party consensus, demonstrating the excellent performance in various practical conditions of our QBA protocol.

We further simulate the consensus rate when using BB84 KGP adopted from the asymmetric coding BB84 QKD~\cite{lo2005Efficient,lim2014concise,yin2020tight}, as shown in Fig.~\ref{Fig5}. Simulations employed the following parameters: detector efficiency of 70\%, dark count rate of $10^{-8}$, misalignment error rate of 0.02, and error correction efficiency of 1.1. Besides different numbers of participants, we additionally calculate the consensus rates for different sizes of the message (1 Kb, 1 Mb, and 100 Mb). The consensus rate exceeds $10^3$ times per second in a 7-party QBA with a 100 Mb message for LEO satellites. These simulations demonstrate the practicality of our QBA protocol, since the experimental requirements of BB84 KGP are simple, and BB84 QKD has been implemented in the satellite-to-ground system recently~\cite{li2025microsatellite}. All the simulation details can be found in Appendix~\ref{Simulation details}.

Note that in our simulations of the circular QBA protocol on satellite-to-ground networks, we employ the commonly used BB84 and DM-CV protocols for key distribution. While adopting more efficient QKD protocols with higher key rates, such as twin-field QKD~\cite{lucamarini2018overcoming,wang2018twin-field} and asynchronous measurement-device-independent QKD~\cite{xie2022breaking,zeng2022mode}, would naturally increase the signature generation rate and the overall consensus rate, this improvement is straightforward.

\section{Discussion and outlook}
The blockchain trilemma describes the challenge of balancing three fundamental aspects of a distributed system: decentralization, scalability, and security. Achieving all three optimally is impossible for classical distributed systems, as improving one aspect often comes at the cost of another. This trade-off also persists in all known QBA protocols, as shown in Table~\ref{Tabel2}. We also compare our protocol with other QBA protocols in terms of the three aspects of the blockchain trilemma to show the practicality of our work. In addition, the blockchain trilemma also remains for our circular QBA due to the relaxation of decentralization. To date, whether a QBA protocol can overcome the blockchain trilemma remains an open question.

Interestingly, our approach demonstrates that breaking the trilemma may not be necessary. Our results indicate that by partially relaxing decentralization to optimize security and scalability, it is possible to design an i.-t, secure multiparty QBA protocol with quadratic communication complexity, a fault tolerance of $N \geq f + 2$, and requiring only $N$ quantum channels, making it suitable for large-scale deployment. 

It is worth noting that the introduction of the CA in our framework is a deliberate design choice to navigate the blockchain trilemma. While the CA is a designated entity, its role is strictly confined to signature verification as a supporting cryptographic infrastructure, rather than acting as a decision-making participant in the consensus process. Furthermore, although the current work assumes a single CA for simplicity, the framework could be extended to multi-CA scenarios in practical network topologies. In such a setup, security against partially corrupted CAs could be achieved through redundancy or threshold-based verification techniques. Exploring the multi-CA architecture for circular QBA remains an interesting open direction for future QBA networks.

This work paves the way for practical, efficient, and scalable QBA networks compatible with current quantum network topologies, and lays the foundation for future quantum blockchain systems capable of delivering i.-t. secure and fault-tolerant decentralized information services, even under the potential threats of quantum computing in the quantum era.

\bigskip
\noindent	
\begin{center}
    \textbf{\large Acknowledgements}
\end{center}
This study was supported by the National Natural Science Foundation of China (No. U25D8016, No. 12522419, No. 12274223), the Program for Innovative Talents, Entrepreneurs in Jiangsu (No. JSSCRC2021484), and the Fundamental Research Funds for the Central Universities and the Research Funds of Renmin University of China (No. 24XNKJ14). We gratefully acknowledged the icons designed by Vecteezy from~\href{https://www.vecteezy.com/}{www.vecteezy.com}.

\appendix

\section{\label{Algorithmic} Algorithmic protocol description}
To make our circular QBA protocol easy to follow and understand, we provide a detailed step-by-step description of its main phases in an algorithmic format as shown in Algorithm~\ref{alg:circular_qba}. 

\begin{algorithm}[h]
    \renewcommand{\algorithmicrequire}{\textbf{Input:}}
    \renewcommand{\algorithmicensure}{\textbf{Output:}}
    \caption{Circular Quantum Byzantine Agreement Protocol}
    \label{alg:circular_qba}
    \begin{algorithmic}[1]
        \REQUIRE $N$ players, security parameter $\lambda$.
        \STATE \textbf{Initialization:} $S \leftarrow \text{commanding general}$; $R_i (i \in \mathbb{Z}_{N-1}^+) \leftarrow \text{lieutenants}$; $CA \leftarrow \text{assistant}$.
        \STATE \textbf{Phase 1: Order Distribution}
        \FOR{$i = 1$ \TO $N-1$}
            \REPEAT
                \STATE $S$ sends $\{m_i;\sigma_i\}$ to $R_i$
                \STATE CA verifies the validity of $\sigma_i$
                \IF{verification fails}
                    \STATE Restart this distribution step
                \ELSE
                    \STATE $R_i$: $O^i \leftarrow \{m_i;\sigma_i\}$
                    \STATE CA records $\sigma_i$ and related quantum keys
                \ENDIF
            \UNTIL{$R_i$ receives valid order}
        \ENDFOR
        \STATE \textbf{Phase 2: Circular Gathering}
        \FORALL{$i \in \mathbb{Z}_{N-1}^+$}
            \STATE $j \leftarrow i$, stipulating $(N-1)+1 \equiv 1$
            \REPEAT
                \IF{$j=i$}
                    \STATE $R_i$: $G^i_{i\rightarrow i+1} \leftarrow O^i =\{m_i;\sigma_i ;\sigma^i_{i\rightarrow i+1} \}$
                \ELSE
                    \STATE $R_j$: $G^i_{j\rightarrow j+1} \leftarrow G^i_{j-1\rightarrow j} \cup O^j \cup \sigma^i_{j\rightarrow j+1}$
                \ENDIF
                \STATE $R_j$ sends $G^i_{j\rightarrow j+1}$ to $R_{j+1}$
                \STATE CA performs verification: (a) $\sigma^i_{j\rightarrow j+1}$ is valid; (b) signatures in $G^i_{j\rightarrow j+1}$ are unchanged; (c) signatures match $\{m_k\}$.
                \IF{any verification fails}
                    \STATE Restart current step
                \ELSE
                    \STATE CA records $\sigma^i_{j\rightarrow j+1}$
                    \STATE $j \leftarrow j+1$
                \ENDIF
            \UNTIL{$R_{i-1}$ sends $G^i_{i-1\rightarrow i}$ to $R_{i}$}
            \STATE $R_i$ extracts $F^i \leftarrow \{ m_1, m_2, \dots, m_{N-1} \}$ from $G^i_{i-1\rightarrow i}$
        \ENDFOR
        \STATE \textbf{Phase 3: Consensus Output}
        \FORALL{$i \in \mathbb{Z}_{N-1}^+$}
            \STATE $R_i$: $m^{R_i} \leftarrow D(F^i) = D(m_1, m_2, \dots, m_{N-1})$
        \ENDFOR
        \ENSURE $(\forall i \in \mathbb{Z}_{N-1}^+)$ Honest $R_i$ outputs $m^{R_i}$.
    \end{algorithmic}
\end{algorithm}

\section{\label{KGP}Key generation protocol } 
The KGP employed in our QBA protocol can be derived from any QKD protocol with two participants. We illustrate the steps of the discrete-modulated continuous-variable (DM-CV) KGP, adapted from the DM-CV QKD with homodyne detection~\cite{liu2021homodyne} and heterodyne detection~\cite{lin2019Asymptotic}, and the asymmetric coding BB84 KGP, adapted from the asymmetric coding BB84 QKD protocol in Refs.~\cite{lo2005Efficient,lim2014concise}. In addition, we give the calculation details of both KGPs in Appendix~\ref{Simulation details}. To complete a round of OTUH-QDS, three participants need to form correlated keys. The KGP involves the signer-forwarder and signer-verifier independently sharing an identical quantum key, and the verifier (CA in our QBA protocol) calculating the bit-wise XOR of the two keys. In the description below, we describe how two participants share the identical quantum keys, and the two participants are denoted as Alice (always CA) and Bob.

\subsection{DM-CV KGP}
Here, we introduce the detailed steps of two DM-CV QKD protocols with homodyne detection and heterodyne detection, respectively. 

\noindent\textbf{(a) Homodyne.}      

\textit{1. Preparation}: In each round, Alice prepares a coherent state $\ket{\alpha e^{i\phi}}$, where $\phi$ is uniformly randomly chosen from $\left\{\frac{\pi}{4}, \frac{3\pi}{4}, \frac{5\pi}{4}, \frac{7\pi}{4}\right\}$. She then sends the state to the receiver, Bob.

\textit{2. Measurement}: Upon receiving the state, Bob uniformly randomly chooses a quadrature from $\left\{\hat{p}, \hat{q}\right\}$ to measure and records the measurement outcome.

\textit{3. Parameter estimation}: After $N$ rounds, they announce the measurement quadratures through the authenticated public channel, and Bob announces a subset of the measurement results in order to evaluate the security, i.e., to compute the secret key rate. If the rate is below zero, the protocol is aborted.

\textit{4. Key generation}: If the protocol is not aborted, suppose Bob obtains $M$ outcomes labeled as $\boldsymbol{Z} = (z_1, \dots, z_k, \dots, z_M)$. Alice labels the four choices as $\left\{00, 10, 11, 01\right\}$. If the quadrature that Bob measures in a given round is $\hat{q}$, Alice sets her key bit $a_k$ as the first bit of the corresponding label. Otherwise, $a_k$ is set as the second bit. As for Bob, if the measurement outcome $z_k \in (-\infty, -\Delta)$, his key bit $b_k$ is set to 1, where $\Delta$ is a non-negative sifting parameter. If $z_k \in (\Delta, \infty)$, $b_k$ is set to 0. Otherwise, $b_k$ is assigned the value $\perp$. They then obtain their key strings $\boldsymbol{A} = (a_1, \dots, a_M)$ and $\boldsymbol{B} = (b_1, \dots, b_M)$.

\textit{5. Error correction and privacy amplification}: After performing error correction and privacy amplification, they obtain the final key strings.

\noindent\textbf{(b) Heterodyne.}

\textit{1. Preparation}: The preparation process is similar. In each round, Alice prepares a coherent state with random phases and sends the state to the receiver Bob.

\textit{2. Measurement}: Instead of measuring a quadrature, Bob directly performs a POVM described by $\{E_{\gamma} = (1/\pi)\ket{\gamma}\bra{\gamma}:\gamma \in \mathcal{C}\}$.

\textit{3. Parameter estimation}: After $N$ rounds, they use a small subset of the data to compute the secret key rate and decide whether to abort the protocol.

\textit{4. Key generation}: Suppose they obtain $M$ measurement outcomes recorded as $\boldsymbol{Z} = (z_1, \dots, z_k, \dots, z_M)$, where each $z_k$ has the form $|z_k|e^{i\theta_k}$. Alice records her bits directly as $\{0,1,2,3\}$ according to her choices. Bob uses two sifting parameters, $\Delta_a$ and $\Delta_p$, separately for the amplitude and phase. The mapping rule from the measurement outcomes $\boldsymbol{Z}$ to his key string $\boldsymbol{B}$ is as follows:

$$
b_k = 
\begin{cases} 
0 & \text{if } \theta_k \in \left[ \Delta_p, \frac{\pi}{2} - \Delta_p\right) \text{ and } |z_k| \geq \Delta_a, \\
1 & \text{if } \theta_k \in \left[\frac{\pi}{2} + \Delta_p, \pi - \Delta_p\right) \text{ and } |z_k| \geq \Delta_a, \\
2 & \text{if } \theta_k \in \left[\pi + \Delta_p, \frac{3\pi}{2} - \Delta_p\right) \text{ and } |z_k| \geq \Delta_a, \\
3 & \text{if } \theta_k \in \left[\frac{3\pi}{2} + \Delta_p, 2\pi - \Delta_p\right) \text{ and } |z_k| \geq \Delta_a, \\
\perp & \text{if do not satisfy any of the conditions}.
\end{cases}
$$

\textit{5. Error correction and privacy amplification}: After the process of error correction and privacy amplification, they obtain the final key strings.

\subsection{BB84 KGP}
Alice and Bob select the $\mathsf{X}$ and $\mathsf{Z}$ bases with probabilities $q_x$ and $1-q_x$, respectively. Key generation employs only those instances where both parties select the $\mathsf{X}$ basis. Phase-randomized laser pulses are transmitted, incorporating a two-decoy setting. Each pulse intensity is randomly chosen from the set $\mathcal{K} = \{\mu_1, \mu_2, \mu_3\}$, with probabilities $p_{\mu_1}$, $p_{\mu_2}$, and $p_{\mu_3} = 1 - p_{\mu_1} - p_{\mu_2}$, subject to the constraints $\mu_1 > \mu_2 + \mu_3$ and $\mu_2 > \mu_3 \ge 0$. The protocol proceeds in five phases described below.

\textit{1. Preparation.} Alice randomly selects a bit value ($y_i$), a basis ($a_i \in \{\mathsf{X}, \mathsf{Z}\}$), and an intensity ($k_i \in \mathcal{K}$). A corresponding laser pulse is then prepared and sent to Bob.

\textit{2. Measurement.} Bob independently selects a basis ($b_i \in \{\mathsf{X}, \mathsf{Z}\}$) and measures the received pulse. The four possible outcomes are $\{0, 1, \emptyset, \perp\}$, where 0 and 1 represent bit values, $\emptyset$ denotes no detection, and $\perp$ indicates a double-click event. Bob records the outcome as $y_i'$, assigning a random bit to $y_i'$ in the case of a double-click event.

\textit{3. Basis reconciliation.} Alice and Bob exchange their basis and intensity choices through an authenticated public channel, identifying the sets $X_k$ and $\mathcal{Z}_k$ for each intensity $k \in \mathcal{K}$. The protocol continues only if $|X_k| \ge n_{X,k}$ and $|\mathcal{Z}_k| \ge n_{\mathcal{Z},k}$ for all $k$. Otherwise, they repeat steps 1 and 2. The total number of pulses sent by Alice is denoted as $N$.

\textit{4. Generation of raw key and error estimation.} A raw key pair $(\mathbf{X}_A, \mathbf{X}_B)$ is generated by sampling $n_{X} = \sum_{k \in \mathcal{K}} n_{X,k}$ elements from $X = \cup_{k \in \mathcal{K}} X_k$. The number of bit errors ($m_{\mathcal{Z},k}$) in the $\mathcal{Z}_k$ sets are then determined, along with the number of vacuum ($s_{X,0}$) and single-photon ($s_{X,1}$) events in $(\mathbf{X}_A, \mathbf{X}_B)$, and the number of phase errors ($c_{X,1}$) in the single-photon events is also calculated. The protocol proceeds only if the phase error rate $\phi_{X} = c_{X,1} / s_{X,1}$ is below the threshold $\phi_{tol}$.

\textit{5. Post-processing.} This phase involves error correction (revealing $\lambda_{EC}$ bits), error verification using two-universal hash functions, and final privacy amplification to generate the secret key pair.

\section{\label{OTUH}OTUH-QDS}
Achieving information-theoretic security in long message signing, the OTUH-QDS framework~\cite{yin2023experimental} utilizes one-time pads for encryption and LFSR-based Toeplitz hashing for signature generation. Here, we first present the construction of LFSR-based Toeplitz hashing, then we detail the two phases of OTUH-QDS and present its security subsequently.

\subsection{LFSR-based Toeplitz hashing}
A hash function acts like a digital fingerprint for a document or message. Even a minor change in the input results in a significantly different hash value. Moreover, it is computationally infeasible to recover the original message from its hash value, meaning there are no known efficient algorithms to do so. This makes hash functions crucial for verifying data integrity and authenticity. LFSR-based Toeplitz hashing~\cite{krawczyk1994lfsr} offers a compelling combination of efficiency and security, which is employed in the OTUH-QDS protocol. We will now explain how to construct an LFSR-based Toeplitz hash function and demonstrate its application in calculating a hash value with length $\mathcal{N}$ for a given message with length $\mathcal{M}$.

The construction of this hash function begins by generating a random irreducible polynomial of order $\mathcal{N}$ in the Galois field GF(2): $p(x)=x^\mathcal{N}+p_{\mathcal{N}-1}x^{\mathcal{N}-1}+\cdots+p_1x+p_0$ (see Ref.~\cite{Shoup1996OnFast} for an efficient method). The polynomial's coefficients, excluding the term of order $\mathcal{N}$, are represented as an $\mathcal{N}$-bit vector $p=(p_{\mathcal{N}-1},p_{\mathcal{N}-2},\cdots,p_1,p_0)^T$, where $T$ means the transpose of vectors. Next, a binary $\mathcal{N}$-bit string obtained from a quantum information process with unconditional security is expressed as an $N$-dimensional vector $s=(s_{\mathcal{N}-1},s_{\mathcal{N}-2},\cdots,s_1,s_0)^T$ with each element is either 0 or 1. In the OTUH-QDS we employed here, this bit-string is the quantum key obtained through a KGP. These two vectors determine an LFSR-based Toeplitz hashing function $H_{s,p}$. We define the following $\mathcal{M}$ vectors:
\begin{equation}
    \begin{aligned}
        &H_1=s=(s_{\mathcal{N}-1},s_{\mathcal{N}-2},\cdots,s_1,s_0)^T,\\
        &H_2=(p* H_1,s_{\mathcal{N}-1},\cdots,s_2,s_1)^T,\\
        &H_3=(p* H_2,p* H_1,\cdots,s_3,s_2)^T,\\
        &\cdots\\
        &H_{\mathcal{M}}=(p* H_{\mathcal{M}-1},p* H_{\mathcal{M}-2},\cdots,p* H_{\mathcal{M}-\mathcal{N}})^T,
    \end{aligned}
\end{equation}
where $*$ denotes the dot product between two vectors. Then, the LFSR-based Toeplitz hashing function $H_{s,p}$ is expressed as a $\mathcal{N}\times \mathcal{M}$ Toeplitz matrix $H_{\mathcal{N}\mathcal{M}}=(H_1, H_2,\cdots, H_\mathcal{M})$.

To calculate the hash value of a message, the message $m$ is first represented as an $\mathcal{M}$-bit vector. This vector is then multiplied by the Toeplitz matrix $H_{\mathcal{N}\mathcal{M}}$, yielding an $\mathcal{N}$-bit vector. Finally, a modulo-2 operation is applied to each element of the resulting vector. The final $\mathcal{N}$-bit vector obtained represents the hash value of the input message, ensuring data integrity and facilitating secure authentication protocols.

\subsection{Detailed steps and signature rate of OTUH-QDS}

\textit{1. Distribution phase.} The initial step involves the establishment of shared secret keys among the signer ($\mathbb{S}$), forwarder ($\mathbb{F}$), and verifier ($\mathbb{V}$) through a KGP. This is achieved through a QKD or quantum secret sharing protocol, resulting in correlated bit strings where the signer's key is the bitwise XOR of the forwarder's and verifier's keys. Specifically, each party possesses three key bit strings, denoted as $\{ X_{\mathbb{S}}, X_{\mathbb{F}}, X_{\mathbb{V}} \}$, $\{ Y_{\mathbb{S}}, Y_{\mathbb{F}}, Y_{\mathbb{V}} \}$, and $\{ Z_{\mathbb{S}}, Z_{\mathbb{F}}, Z_{\mathbb{V}} \}$. Each string, comprised of $\mathcal{N}$ bits, fulfills the relations $X_{\mathbb{S}}=X_{\mathbb{F}}\oplus X_{\mathbb{V}}$, $Y_{\mathbb{S}}=Y_{\mathbb{F}}\oplus Y_{\mathbb{V}}$, and $Z_{\mathbb{S}}=Z_{\mathbb{F}}\oplus Z_{\mathbb{V}}$.

\textit{2. Messaging phase.} This process begins with the signer generating an LFSR-based Toeplitz hash function~\cite{krawczyk1994lfsr} $H_{X_{\mathbb{S}},p_{\mathbb{S}}}$, which is determined by the secret key $X_{\mathbb{S}}$ and a local random string $p_{\mathbb{S}}$ which corresponds to the coefficients of a irreducible polynomial. The message to be signed is denoted as $\mathrm{Mes}$ with length $\mathcal{M}$. The signer calculates the $\mathcal{N}$-bit digest value $\mathrm{Dig}=H_{X_{\mathbb{S}},p_{\mathbb{S}}}(\mathrm{Mes})$ and encrypts it with $Y_{\mathbb{S}}$ to the signature for this message $\mathrm{Sig}=\mathrm{Dig}\oplus Y_{\mathbb{S}}$. The signer also encrypts $p_{\mathbb{S}}$ with $Z_{\mathbb{S}}$ to $p=p_{\mathbb{S}}\oplus Z_{\mathbb{S}}$ and subsequently transmits $\{\mathrm{Mes}, \mathrm{Sig}, p \}$ through an authenticated channel to the forwarder.

After receiving $\{\mathrm{Mes}, \mathrm{Sig}, p \}$, the forwarder transmits this data, along with key bit strings $\{ X_{\mathbb{F}}, Y_{\mathbb{F}}, Z_{\mathbb{F}} \}$, to the verifier through an authenticated channel. Subsequently, the verifier forwards key bit strings $\{ X_{\mathbb{V}}, Y_{\mathbb{V}}, Z_{\mathbb{V}} \}$ to the forwarder via the same channel. The forwarder then computes three new key bit strings: $K_{X_{\mathbb{F}}} = X_{\mathbb{F}} \oplus X_{\mathbb{V}}$, $K_{Y_{\mathbb{F}}} = Y_{\mathbb{F}} \oplus Y_{\mathbb{V}}$, and $K_{Z_{\mathbb{F}}} = Z_{\mathbb{F}} \oplus Z_{\mathbb{V}}$.  Using $K_{Y_{\mathbb{F}}}$, the forwarder decrypts the signature to obtain the expected digest value, $\mathrm{Dig}_{\mathrm{f1}} = \mathrm{Sig} \oplus K_{Y_{\mathbb{F}}}$. Similarly, $K_{Z_{\mathbb{F}}}$ is used to decrypt $p$, revealing the string $p_{\mathbb{F}} = p \oplus K_{Z_{\mathbb{F}}}$. Next, the forwarder employs $K_{X_{\mathbb{F}}}$ and $p_{\mathbb{F}}$ to construct an LFSR-based Toeplitz hash function, which is then applied to the message $\mathrm{Mes}$. This operation yields the actual digest value, $\mathrm{Dig}_{\mathrm{f2}}$. Finally, the forwarder compares the expected and actual digest values. If $\mathrm{Dig}_{\mathrm{f1}}$ and $\mathrm{Dig}_{\mathrm{f2}}$ are identical, the signature is deemed valid; otherwise, it is considered invalid. The forwarder then announces the result.

The verifier also performs an authentication process after receiving $\{\mathrm{Mes}, \mathrm{Sig}, \textit{p} \}$ and the forwarder's keys $\{ X_{\mathbb{F}}, Y_{\mathbb{F}}, Z_{\mathbb{F}} \}$ if the forwarder declares acceptance. The verifier calculates keys $K_{X_{\mathbb{V}}} = X_{\mathbb{V}} \oplus X_{\mathbb{F}}$, $K_{Y_{\mathbb{V}}} = Y_{\mathbb{V}} \oplus Y_{\mathbb{F}}$, and $K_{Z_{\mathbb{V}}} = Z_{\mathbb{V}} \oplus Z_{\mathbb{F}}$. Similar to the forwarder's actions, the verifier utilizes the $Y_{\mathbb{V}}$ and $Z_{\mathbb{V}}$ to decrypt $\mathrm{Sig}$ and $p$ respectively, and obtains the expected digest value and a string $p_{\mathbb{V}}$. With $K_{X_{\mathbb{V}}}$ and $p_{\mathbb{V}}$, the verifier constructs an LFSR-based Toeplitz hash function and computes the actual digest value of the message. Finally, the verifier compares the expected and actual digest values to determine the validity of the signature, authenticating the signature if they match and rejecting it otherwise. We note that in our QBA protocol, the verifier is referred to as the assistant CA. CA needs to inform the forwarder, i.e., a lieutenant, of the result of verification, thereby ensuring an independent verification of the message's authenticity and integrity.

The signature rate (SR) defined as the number of secure signatures that can be generated and checked per second, is determined by~\cite{yin2023experimental}
\begin{equation}
    SR = KR / 3n,
\end{equation}
where $KR$ is the quantum key rate between two participants and $n$ is the length of the signature, reflecting that signing an $n$-bit signature using OTUH-QDS with LFSR-based Toeplitz hashing requires $3n$ bits of quantum keys.

\subsection{\label{Security of OTUH-QDS}Security of OTUH-QDS}
The security of QDS relies on two essential properties: unforgeability and non-repudiation. This appendix examines the security foundations of the OTUH-QDS protocol, demonstrating its security against both classical and quantum attacks. 

In a KGP process, successful privacy amplification eliminates information leakage during key generation. Consequently, an attacker's strategy for guessing the $\mathcal{N}$-bit key is reduced to randomly generating an $\mathcal{N}$-bit string and comparing it to the true key. This trivial method yields a success probability of $2^{-\mathcal{N}}$. With this relation, we can proceed to investigate potential attack strategies against almost XOR universal$_2$ (AXU) hashing, specifically focusing on the LFSR-based Toeplitz hashing method described previously. A naive approach for an attacker involves randomly generating two bit strings, $t$ of length $\mathcal{M}$ and $h$ of length $\mathcal{N}$. If, by chance, $h$ coincides with the hash value of $t$ under the employed hash function, the attacker can intercept the transmitted message $\mathrm{Mes}$ and signature $\mathrm{Sig}$ and manipulate them to $\mathrm{Mes}^{\prime} = \mathrm{Mes} \oplus t$ and $\mathrm{Sig}^{\prime} = \mathrm{Sig} \oplus h$. Due to the properties of the hash function, the recipient would unknowingly accept the tampered message and signature pair as valid. The success probability of this approach is only $2^{-\mathcal{N}}$. However, the inherent information leakage during key generation presents the attacker with two potential avenues for compromising the system: attempting to guess the key $Key$ that defines the hashing function based on the information leakage, or endeavouring to derive this key from the intercepted signature. As demonstrated in Ref.~\cite{li2023one}, a specific property of the LFSR-based Toeplitz hash function restricts the number of viable guesses for $Key$ to be within the range of $\mathcal{M}/\mathcal{N}$. Furthermore, considering the attacker's awareness of the chosen polynomial's irreducible nature, the success probability of this guessing strategy can be expressed as:
\begin{equation}
    p_{\rm{att}} = \frac{\mathcal{M}}{\mathcal{N}} P [\text{guess}\ Key\ |\ p_g\ \text{is irreducible}],
\end{equation}
where $P[ \cdot ]$ denotes the probability of an event and $p_g$ represents the attacker's chosen irreducible polynomial. According to the characteristics of GF(2), there more than $2^{\mathcal{N}-1}/\mathcal{N}$ irreducible polynomials of order $\mathcal{N}$. Thus, we have $P [\text{is irreducible}] \ge \frac{2^{\mathcal{N}-1}/\mathcal{N}}{2^\mathcal{N}} = 1/2\mathcal{N}$. Leveraging Bayes' theorem, we can establish an upper bound for $p_{\rm{att}}$:
\begin{equation}
    \begin{aligned}
        p_{\rm{att}} &= \frac{\mathcal{M}}{\mathcal{N}} \cdot \frac{P[\text{guess}\ Key]}{P[p_g\ \text{is irreducible}]} \\
        &= \frac{\mathcal{M}}{\mathcal{N}} \cdot \frac{P(Key|\mathcal{B})}{P[p_g\ \text{is irreducible}]}\\
        &\le \frac{\mathcal{M}}{\mathcal{N}} \cdot \frac{2^{-\mathcal{N}}}{1/2\mathcal{N}}=\mathcal{M}\cdot2^{1-\mathcal{N}}=\varepsilon_{\rm{U}}(\mathcal{M}, \mathcal{N}).
    \end{aligned}
\end{equation}
This upper bound is a function of the message and signature lengths, as $\mathcal{N}$ is dependent on $\mathcal{N}$. It is obvious that the probability of the attacker successfully guessing the other two keys, $Y$ and $Z$, which are employed for encryption, is also constrained by $\varepsilon_{\rm{U}}(\mathcal{M}, \mathcal{N})$. Furthermore, the alternative strategy of deriving the key $X$ from the intercepted signature also necessitates correctly guessing the key responsible for encrypting the coefficients of the irreducible polynomial. Consequently, the success probability associated with this approach remains confined by the same upper bound, $\varepsilon_{\rm{U}}(\mathcal{M}, \mathcal{N})$. Overall, the optimal attack strategy against LFSR-based Toeplitz hashing focuses on guessing the key string that either determines the hashing function or encrypts the polynomial used within it. Consequently, regardless of the specific approach employed by the attacker, the upper bound of the failure probability for authentication schemes remains $\varepsilon_{\rm{U}}(\mathcal{M}, \mathcal{N}) =\mathcal{M}\cdot2^{1-\mathcal{N}}$.

Unforgeability ensures that an attacker cannot create a valid signature for a tampered message. A successful forgery would occur if the attacker could find a tampered message that, after one-time pad decryption and one-time AXU hashing, yields the same hash value as the original message, thereby leading to its acceptance. However, as previously analyzed, even considering potential information leakage during key generation, the probability of a successful forgery remains negligible. Specifically, the probability of successful forgery $\varepsilon_{\rm{for}}$ is directly equivalent to the failure probability of the hashing function, leading to the expression:
\begin{equation}
    \varepsilon_{\rm{for}}(\mathcal{M}, \mathcal{N})=\varepsilon_{\rm{U}}(\mathcal{M}, \mathcal{N}) =\mathcal{M}\cdot2^{1-\mathcal{N}}.
\end{equation}

Non-repudiation guarantees that the signer cannot deny having signed a message or claim forgery. This QDS protocol is inherently secure against repudiation as the forwarder and verifier can recover the keys used by the signer through the authenticated classical channel, which is impervious to the signer's influence. The forwarder and verifier can then independently obtain the same actual and expected hash results, leading to the same decision. Therefore, neglecting negligible errors inherent in classical communication channels, the probability of successful repudiation is effectively zero.

\section{\label{securityAnalysis}Security analysis of circular QBA}
A Byzantine agreement is capable of reaching a consensus when and only when it satisfies the two \textbf{IC} Byzantine conditions. Our security analysis demonstrates that every lieutenant receives an identical set of input messages ($\textbf{IC}_1$). Furthermore, if the commanding general is loyal, the output will correspond to the commanding general's order ($\textbf{IC}_2$). Unforgeability of QDS ensures that no one can tamper with the message, and non-repudiation ensures that the signer cannot deny the fact that he signs the message. These properties ensure the security of our QBA protocol.

We assume that each message $m_j$ has a length of $m$ bits, while each signature $\sigma_j$ or $\sigma^i_{j \rightarrow j+1}$ ($j \in \mathbb{Z}_{N-1}^+$) has a length of $n$ bits. Assuming a total of $N$ players, with $f$ of them being dishonest, we let $\tau_h$ and $\tau_d$ represent the set of indices corresponding to the honest and dishonest lieutenants respectively: 
\begin{equation}
\begin{aligned}
    &\tau_h = \{ j_h \in \mathbb{Z}_{N-1}^+ ~|~ R_{j_h} \text{ is honest}\},\\
    &\tau_d = \{ j_d \in \mathbb{Z}_{N-1}^+ ~|~ R_{j_d} \text{ is dishonest}\}.
\end{aligned}
\end{equation}
The number of honest and dishonest lieutenants, denoted as $|\tau_h| = N_h$ and $|\tau_d| = N_d$ respectively, are dependent on whether the commanding general is loyal or disloyal. Thus, we will consider two scenarios based on the loyalty of the commanding general.

The security analysis is constructed as follows. In Appendix~\ref{security of CC}, we analyze the security of circular gathering phase which is the building block of our circular QBA. In Appendices~\ref{loyal commanding general} and~\ref{disloyal commanding general} we then prove the security of our protocol in scenarios where the commanding general is loyal and disloyal respectively. Finally, we analyze the overall security and the fault-tolerance ability of our circualr QBA in Appendix~\ref{QBA security}.

\begin{figure}[t]
    \centering
    \includegraphics[width=\linewidth]{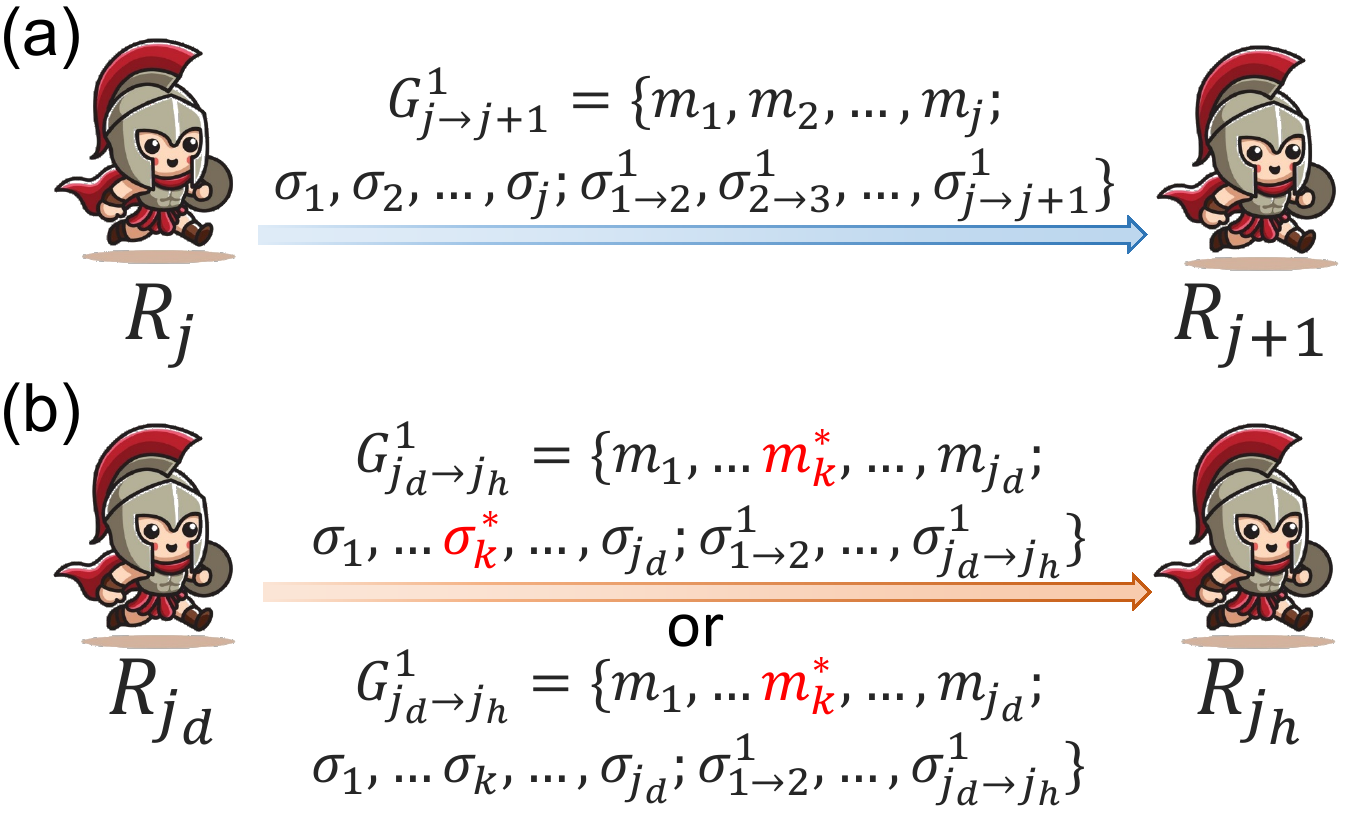}
    \caption{(a) The information delivery step between $R_{j}$ and $R_{j+1}$ within $R_1$'s circular gathering. The information conveyed from $R_i$ to $R_{j+1}$ comprises $j$ message segments ($\{m_1, m_2, \cdots, m_i\}$) along with $2j-1$ signature elements ($\{\sigma_1, \sigma_2, \cdots, \sigma_j \}$ and $\{ \sigma^{1}_{1 \rightarrow 2}, \sigma^{1}_{2 \rightarrow 3}, \cdots, \sigma^{1}_{j-1 \rightarrow j} \}$), resulting in an overall length expressed as $L_{j} = jm + (2j-1)n$. (b) A disloyal lieutenant $R_{j_d}$, acting as a signer, may attempt to substitute either the original message $m_k$ with a forged message $m_k^*$, or the message-signature pair $\{m_k, \sigma_k\}$ with a forged pair $\{m_k^*, \sigma_k^*\}$, where $k \le j$, and sends to a loyal lieutenant $R_{j_h}$. Both are prevented by CA, who retains a record of the original signatures and corresponding quantum keys in the order distribution phase. Malicious signers may also change the signatures $\{ \sigma^1_{1 \rightarrow 1},\cdots,\sigma^1_{j_d-1 \rightarrow j_d} \}$, however, CA also compares them with previous records, preventing this way of cheating. Icons Copyright 2026 Vecteezy.
}
    \label{FigS1}
\end{figure}

\subsection{\label{security of CC}Security of circular gathering phase with malicious lieutenants}
The rounds of circular gathering started by disloyal lieutenant need not to be considered since what disloyal lieutenants obtain and how they generate the output is irrelevant to two \textbf{IC} conditions. Assuming $R_1$ is loyal, the security analysis of $R_1$'s circular gathering serves as a representative example, as the analysis readily generalizes to the circular gathering initiated by other lieutenants due to their symmetrical roles. Figure~\ref{FigS1}(a) highlights a single step where lieutenant $R_j$ transmits messages and signatures to $R_{j+1}$ ($j \in \mathbb{Z}_{N-1}^+$) in the circular gathering of $R_1$.

Within a round initiated by an honest lieutenant, each disloyal lieutenant has one opportunity to sign and one to forward information. When signing, a disloyal lieutenant $R_{j_d}$ could attempt one of two forgery strategies when transmitting messages to an honest lieutenant $R_{j_h}$: (1) substituting the message-signature pair $\{m_k; \sigma_k\}$ with a forged pair $\{m_k^*; \sigma_k^*\}$ ($k \le j$), where $\sigma_k^*$ is a forged signature corresponding to $m_k^*$, or (2) substituting only the original message $m_k$ with a forged message $m_k^*$. These two strategies are illustrated in Fig.~\ref{FigS1}(b). The assistant CA, by preserving the original signatures and quantum keys recorded during the order distribution phase, effectively obstructs both strategies. CA prevents the first by comparing all signatures against the original records. The second is prevented by verifying the consistency between $m_k^*$ and $\sigma_k$ using one-time universal hashing via quantum keys recorded during order distribution. While the messages themselves appear unchanged, it's important to note that the process $R_{j_d}$ may change some of the signatures generated during the circular gathering phase, specifically those within the set $\{\sigma^1_{1 \rightarrow 2}, \dots, \sigma^1_{j_d-1 \rightarrow j_d}\}$. This way of cheating is also prevented since CA will compare them with the record of the previous OTUH-QDS in this circular gathering.

Therefore, the key challenge lies in analyzing the potential for forgery when a disloyal lieutenant acts as a forwarder. We first analyze an information delivery step. As shown in Fig.~\ref{FigS1}(a), $R_j$ signs the total information, which are $j$ pieces of messages ($\{m_1, m_2, \cdots, m_j\}$) and $2j-1$ pieces of signatures ($\{\sigma_1,\sigma_2,\cdots,\sigma_j \}$ and $\{ \sigma^{1}_{1 \rightarrow 2}, \sigma^{1}_{2 \rightarrow 3},\cdots,\sigma^{1}_{j-1 \rightarrow j} \}$), with the signature $\sigma^{1}_{j \rightarrow j+1}$, and then sends all of them to $R_{j+1}$. The length of total transmitted information from $R_j$ to $R_{j+1}$ is $L_{j} = jm + (2j-1)n$, and the length of signature $\sigma^{1}_{j \rightarrow j+1}$ is $n$. So the probability of successful forgery in this QDS process is then expressed as:
\begin{equation}
    \varepsilon^{\rm{D}}_{j} = \varepsilon_{\rm{for}}(L_{j}, n).
\end{equation}
According to Appendix~\ref{Security of OTUH-QDS}, as $L_j$ increases, the probability of successfully forging increases since the signature's length is fixed to $n$. This will be used to provide an upper bound for the failure probability of a circular gathering round. 

A complete round of circular gathering involves $N-1$ information delivery steps and includes $N-1$ lieutenants. Since only malicious lieutenants would attempt forgery, and to guarantee that $R_1$ receives the correct messages, no QDS forgery should occur in any of the information delivery steps. Therefore, the probability of a successful circular gathering initiated by $R_1$ is calculated as follows 
\begin{equation}
    P_{\rm{c}}^{R_1}(\tau_d) = \prod_{j_d \in \tau_d} (1-\varepsilon^{\rm{D}}_{j_d}) = \prod_{j_d \in \tau_d} \left [1-\varepsilon_{\rm{for}}(L_{j_d}, n) \right ].
\end{equation}

Our goal is to guarantee that all loyal lieutenants receive the same set of messages. Therefore, we focus on the circular gatherings initiated by loyal lieutenants, denoted by $R_{j_h}$ ($j_h \in \tau_h$). All such gatherings must be successful. Moreover, $P_{\rm{c}}^{R_1}(\tau_d)$ is influenced by the permutations of the dishonest lieutenants within the gathering circle. Specifically, because longer messages are easier to forge, a dishonest lieutenant positioned later in the sequence poses a greater security risk. To establish a rigorous upper bound on the probability of failure, we consider the worst-case scenario for $R_1$, where $P_{\rm{c}}^{R_1}(\tau_d)$ is minimized. We then use this minimized value as the probability for the success of circular gatherings initiated by all loyal lieutenants. This approach is conservative, as the worst-case scenario for $R_1$ might not be the worst case for other loyal lieutenants. This simplification, however, allows for a stricter security analysis. Consequently, the probability of all circular gatherings to be successful is:
\begin{equation}
\begin{aligned}
     P_{\rm{c}}(\tau_d) &= \prod_{j_h \in \tau_h} P_{\rm{c}}^{R_{j_h}}(\tau_d) \ge \left [ \inf_{\substack{\tau_d \subset \mathbb{Z}_{N-1}^+\\|\tau_d|=N_d}} P_{\rm{c}}^{R_1}(\tau_d)  \right ]^{N_h} \\
     &= \left \{ \inf_{\substack{\tau_d \subset \mathbb{Z}_{N-1}^+\\|\tau_d|=N_d}} \prod_{j_d \in \tau_d} \left [1-\varepsilon_{\rm{for}}(L_{j_d}, n) \right ]  \right \}^{N_h}\\
     &\ge \left \{ \inf_{\substack{\tau_d \subset \mathbb{Z}_{N-1}^+\\|\tau_d|=N_d}} \left [ 1-\sum_{j_d \in \tau_d} \varepsilon_{\rm{for}}(L_{j_d},n) \right ]  \right \}^{N_h} \\
     &=\left[ 1- \sup_{\substack{\tau_d \subset \mathbb{Z}_{N-1}^+\\|\tau_d|=N_d}} \sum_{j_d \in \tau_d} \varepsilon_{\rm{for}}(L_{j_d},n)  \right]^{N_h},
\end{aligned}
\end{equation}
where the infimum ($\inf$) and supremum ($\sup$) are taken over all possible sets $\tau_d$ (a subset of $\mathbb{Z}_{N-1}^+$ with $N_d$ elements), representing all possible sequences dishonest lieutenants. The second scaling holds because $\varepsilon_{\rm{for}}(L_{j},n)$ is exceedingly small for all $j$. Therefore, we can disregard terms involving the product of two or more of these small probabilities, as their contribution to the succeeding probability is negligible. 

For the convenience of numerical simulations and the setting of experimental parameters, we can introduce further scaling. The longest message to be signed in a circular gathering has a length of $L_{N-1} = (N-1)m+(2N-3)n$, resulting in the largest probability of forgery $\varepsilon_{\rm{for}}(L_{N-1},n)$. So we have $\varepsilon_{\rm{for}}(L_{j_d},n) \le \varepsilon_{\rm{for}}(L_{N-1},n)$ ($\forall j_d \in \tau_d$). We introduce the scaling as follows:
\begin{equation}
    \sup_{\substack{\tau_d \subset \mathbb{Z}_{N-1}^+\\|\tau_d|=N_d}} \sum_{j_d \in \tau_d} \varepsilon_{\rm{for}}(L_{j_d},n) < N_d   \cdot \varepsilon_{\rm{for}}(L_{N-1},n)
\end{equation}
and the probability of all circular gathering to be successful is then (Note that $N_h+N_d = N-1$):
\begin{equation}
\begin{aligned}
    P_{\rm{c}}(\tau_d) &\ge \left [ 1-N_d   \cdot \varepsilon_{\rm{for}}(L_{N-1},n) \right ]^{N_h}\\
    &\ge 1-N_d   N_h   \cdot \varepsilon_{\rm{for}}(L_{N-1},n)\\ 
    &=1-N_d\left (N-N_d-1\right ) \cdot \varepsilon_{\rm{for}}(L_{N-1},n)  \\
    &=P_{\rm{c}}^{\rm{inf}}(N_d) = 1-\varepsilon_{\rm{c}}^{\sup}(N_d).
\end{aligned}
\end{equation}
The second scaling also holds since $\varepsilon_{\rm{for}}(L_{N-1},n) \ll 1$. Here, $P_{\rm{c}}^{\rm{inf}}(N_d)$ represents the lower bound on the probability of successful circular gatherings, while $\varepsilon_{\rm{c}}^{\sup}(N_d)=N_d(N-N_d-1 )     \cdot \varepsilon_{\rm{for}}(L_{N-1},n)$ represents the upper bound on the probability of failure in any circular gathering. Both are functions of $N_d$, the number of disloyal lieutenants.

\begin{figure*}
    \centering
    \includegraphics[width=\linewidth]{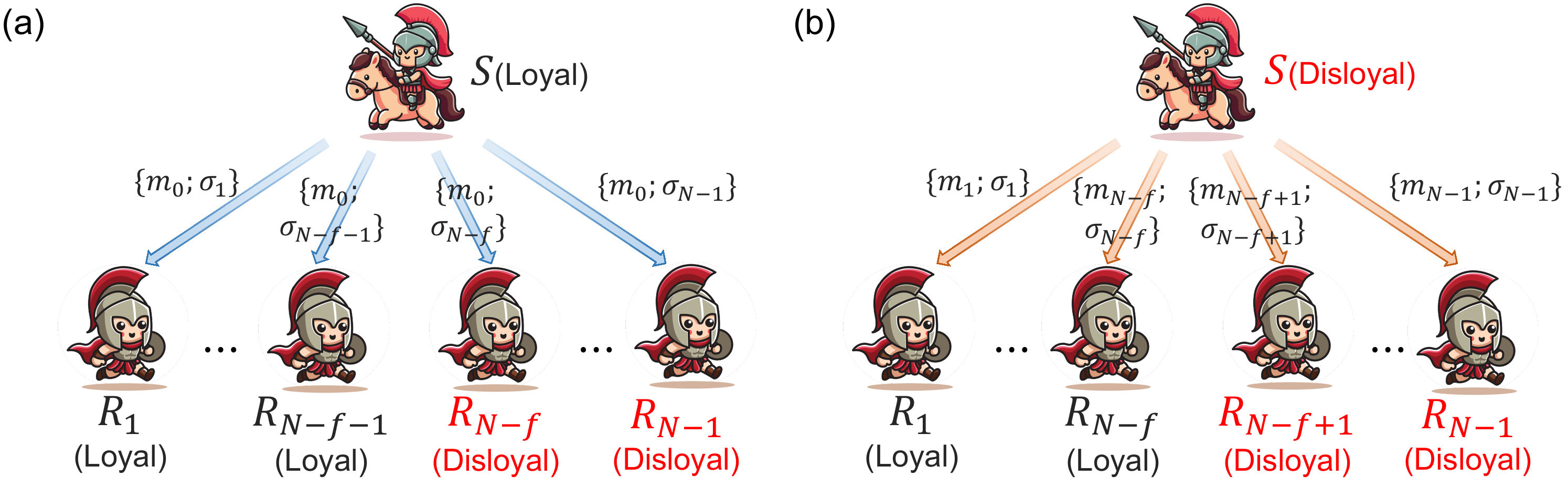}
    \caption{The order distribution phases of two cases. (a) Case \Rmnum{1}: Loyal commanding general. In this case, the same message is distributed. The disloyal lieutenants will try to forge the correct messages. Loyal $S$ broadcasts the same message $m_0$, signed with individual signatures $\sigma_j$ ($j \in \mathbb{Z}_{N-1}^+$), to each lieutenant.  (b) Case \Rmnum{2}: Disloyal commanding general. With a disloyal commanding general and $f-1$ disloyal lieutenants, the order distribution phase involves the broadcast of $N-1$ distinct messages. The goal of malicious players is to maximize uncertainty, which is achieved by the distribution of these distinct messages.
}
    \label{FigS2}
\end{figure*}

\subsection{\label{loyal commanding general}Case \Rmnum{1}: Loyal commanding general}

We first analyze the scenario where the commanding general $S$ is loyal and the number of disloyal lieutenants is $f$. In this case, we have $N_d = f$ and $N_h = N-f-1$. During the order distribution phase shown in Fig.~\ref{FigS2}(a), the loyal $S$ signs the same message $m_0$ with signature $\sigma_i$ ($i \in \mathbb{Z}_{N-1}^+$) and transmits them to lieutenants, with CA acting as the verifier in OTUH-QDS. If the QDS between $S$, $R_i$, and CA is successful, the probability of successful forgery is bounded by $\varepsilon_{\rm{o}}^{S \rightarrow R_i} = \varepsilon_{\rm{for}}(m, n)$. In the order distribution phase, forgery can only occur with a very low probability in any QDS with a disloyal lieutenant since $S$ and CA are honest. Therefore, the probability of a successful order distribution phase is:
\begin{equation}
\begin{aligned}
    P_{\rm{\Rmnum{1}o}} &= \prod_{j_d \in \tau_d} (1-\varepsilon_{\rm{o}}^{S \rightarrow R_{j_d}}) = [1-\varepsilon_{\rm{for}}(m, n)]^{N_d}\\
    &\ge 1-N_d   \cdot \varepsilon_{\rm{for}}(m, n) = 1-f   \cdot \varepsilon_{\rm{for}}(m, n)\\
    &=1-\varepsilon_{\rm{\Rmnum{1}o}}.
\end{aligned}
\end{equation}
Here, $\varepsilon_{\rm{\Rmnum{1}o}} = f   \cdot \varepsilon_{\rm{for}}(m, n)$ means the probability of failure in the order distribution phase in case \Rmnum{1}.

In the circular gathering phase, the protocol must ensure that all $N-f-1$ loyal lieutenants receive the same set of messages, even if the $f$ disloyal lieutenants attempt to disrupt the process. In Case \Rmnum{1}, the number of disloyal lieutenants is $N_d = f$. As analysed in Appendix~\ref{security of CC}, the probability of a successful circular gathering phase is expressed by:
\begin{equation}
\begin{aligned}
    P_{\rm{\Rmnum{1}c}} \ge P_{\rm{c}}^{\rm{inf}}(f) &= 1-f\left (N-f-1\right )  \cdot \varepsilon_{\rm{for}}(L_{N-1},n)\\
    &=1-\varepsilon_{\rm{\Rmnum{1}c}}
\end{aligned}
\end{equation}
where $\varepsilon_{\rm{\Rmnum{1}c}} = f\left (N-f-1\right ) \cdot \varepsilon_{\rm{for}}(L_{N-1},n)$ is the probability of failure in the circular gathering phase in case \Rmnum{1}.

As a result, in case \Rmnum{1} where the commanding general is honest, the probability of a successful operation of the whole protocol is:
\begin{equation}
\begin{aligned}
    P_{\rm{\Rmnum{1}}} = P_{\rm{\Rmnum{1}o}} \cdot P_{\rm{\Rmnum{1}c}}
        &=(1-\varepsilon_{\rm{\Rmnum{1}o}})(1-\varepsilon_{\rm{\Rmnum{1}c}}) \ge 1-\left (\varepsilon_{\rm{\Rmnum{1}o}}+\varepsilon_{\rm{\Rmnum{1}c}} \right ),
\end{aligned}
\end{equation}
since the failure probabilities $\varepsilon_{\rm{\Rmnum{1}o}}$ and $\varepsilon_{\rm{\Rmnum{1}c}}$ are both small. Therefore, the security parameter, which stands for the failure probability of case \Rmnum{1} is then:
\begin{equation}
\begin{aligned}
    \varepsilon_{\rm{\Rmnum{1}}} &= \varepsilon_{\rm{\Rmnum{1}o}}+\varepsilon_{\rm{\Rmnum{1}c}}\\
    &= f   \cdot \varepsilon_{\rm{for}}(m, n) + f\left (N-f-1\right ) \cdot \varepsilon_{\rm{for}}(L_{N-1},n)\\
    &= f\left [  \varepsilon_{\rm{for}}(m, n) + \left (N-f-1\right ) \cdot \varepsilon_{\rm{for}}(L_{N-1},n) \right ]
\end{aligned}
\end{equation}

\subsection{\label{disloyal commanding general}Case \Rmnum{2}: Disloyal commanding general}
In this scenario, there are $N_d = f-1$ disloyal lieutenants and $N_h = N-f$ honest lieutenants. The order distribution phase is depicted in Fig.~\ref{FigS2}(b). A disloyal commanding general aims to disrupt the protocol and mislead the honest lieutenants, sending $N-1$ different messages during the order distribution phase. In this context, forging QDS is irrelevant, as the malicious actors' objective is simply to create maximal confusion. Forging a distributed message by substituting it with another is functionally equivalent to the disloyal commanding general having initially transmitted the substituted message.

The analysis of the circular gathering phase is similar to the previous case. Given $N_d = f-1$, as per the analysis in Appendix~\ref{security of CC}, the probability of a successful circular gathering phase of case \Rmnum{2} $P_{\rm{\Rmnum{2}c}}$, and therefore the QBA protocol's success probability in this scenario $P_{\rm{\Rmnum{2}}}$, is:
\begin{equation}
    \begin{aligned}
        P_{\rm{\Rmnum{2}}} &= P_{\rm{\Rmnum{2}c}} \ge P_{\rm{c}}^{\rm{inf}}(f-1)\\
        &=1-(f-1)(N-f)\cdot \varepsilon_{\rm{for}}(L_{N-1},n)\\
        &=1-\varepsilon_{\rm{\Rmnum{2}}}
    \end{aligned}
\end{equation}
where 
\begin{equation}
    \varepsilon_{\rm{\Rmnum{2}}} = (f-1)(N-f)\cdot \varepsilon_{\rm{for}}(L_{N-1},n),
\end{equation}
representing the failure probability of case \Rmnum{2}.

\subsection{\label{QBA security}Overall security}
Obviously, the two scenarios where the commanding general is loyal and disloyal are complementary. To provide a stricter bound, the security parameter of our QBA scheme, denoting the failure probability of protocol, is expressed as follows:
\begin{equation}
    \begin{aligned}
        \varepsilon_{\rm{QBA}} &= \max \{\varepsilon_{\rm{\Rmnum{1}}}, \varepsilon_{\rm{\Rmnum{2}}} \}\\
        &=\max \{ f\left [  \varepsilon_{\rm{for}}(m, n) + \left (N-f-1\right ) \cdot \varepsilon_{\rm{for}}(L_{N-1},n) \right ], \\
        &(f-1)(N-f)\cdot \varepsilon_{\rm{for}}(L_{N-1},n)\}.
    \end{aligned}
\end{equation}

\section{Simulation details\label{Simulation details}} 
In this part, we give the simulation details of circular QBA networks using different QKG protocols for the OTUH-QDS process.

\subsection{Satellite-to-Ground Channel}
Here, we first present the numerical analysis for the satellite-to-ground link, focusing on the factors affecting laser signal transmission, namely attenuation, beam distortion, and deflection. This model is also employed in Ref.~\cite{Li2024Discrete}. Within this framework, the receiver detects incoming photons, considering both the channel's transparency and environmental mode. The influence of the environmental modes, which cause variations in measurement outcomes, is quantified by the excess noise parameter
\begin{equation}
\xi = \frac{(\Delta q_{\rm obs})^2}{(\Delta q_{\rm vac})^2}-1,
\end{equation}
where $(\Delta q_{\rm obs})^2$ and $(\Delta q_{\rm vac})^2$ stand for the variances of the $q$ quadrature for the observed states and the vacuum states, respectively. 

Considering the signal loss due to photon absorption within the atmosphere, this effect is characterized by the atmospheric transmittance due to extinction, which is denoted by the Beer-Lambert equation \cite{vasylyev2019satellite}
\begin{equation}
    \eta_{\rm ext}(h,\theta) = \exp\left\{-\int_{0}^{z} \alpha[h(y,\theta)] \,dy\right\}.
\end{equation}
Here, $\alpha(h) = \alpha_0 e^{-h/\tilde{h}}$ denotes the altitude-dependent extinction coefficient, where $h$ is the satellite's altitude and $\tilde{h} = 6600~\rm m$ is a scale parameter. Considering both aerosol and molecular scattering, the extinction coefficient at sea level, $\alpha_0$, is $4\times10^{-7}~\rm m^{-1}$ for laser with $1550~\rm nm$-wavelength, and $5\times10^{-6}~\rm m^{-1}$ for laser with $800~\rm nm$-wavelength.

Beyond simple absorption, the non-uniform structure of the atmosphere in both spatial and temporal dimensions leads to phenomena such as beam broadening and wandering~\cite{gunthner2017quantum,vasylyev2019satellite}. For short exposure times, spatial turbulence distorts the beam, causing a broadened beam waist ($W_{\rm ST}$) and a shifted beam center \cite{dios2004scintillation}. A common approximation for the beam waist under turbulence is denoted by \cite{yura1973short}:
\begin{equation}
    W_{\rm ST}^2(z) = W_0^2 \left(1+\frac{z^2}{Z_0^2}\right) + \frac{35.28z^2}{k^2 r_s^2} \left[1-0.26 \left(\frac{r_s}{W_0}\right)^{1/3}\right]^2,
\end{equation}
where $W_0$ is the initial beam waist, $z$ is the propagation distance, $Z_0= kW_0^2/2$ is the Rayleigh distance, $k$ is the wavenumber, and $r_s$ is a measure of turbulence intensity. The turbulence intensity factor $r_s$ is determined by the refractive index structure parameter $C_n^2(h)$
\begin{equation}
    r_s = \left[ 0.42k^2\int_0^L C_n^2(h(z,\theta_z))\left(\frac{L-z}{L}\right)^{5/3} dz\right]^{-3/5},
\end{equation}
where $L$ is the total path length and $C_n^2(h)$ follows the Hufnagel-Valley model \cite{andrews2005laser}
\begin{equation}
\begin{split}
    C_n^2(h) &= 8.1481\times10^{-56} v^2 h^{10} e^{-h/1000} +\\
 &2.7\times10^{-16} e^{-h/1500} + C_0 e^{-h/100}.
\end{split}
\end{equation}
Here, $v$ means the wind speed (21 m/s in our simulation), $h$ is altitude, and $C_0=9.6\times 10^{-14} \rm m^{-2/3}$ is a constant~\cite{qing2016use}. 

Over longer periods, the deviation $r=\sqrt{x^2+y^2}$ of the beam center from the intended path is described by a random distribution
\begin{equation}
    P(x,y)= \frac{1}{2\pi \sigma_r^2} \exp\left(-\frac{x^2+y^2}{2\sigma_r^2} \right),
\end{equation}
The variance $\sigma_r^2$ accounts for both atmospheric effects and pointing errors, defined as $\sigma_r^2 = (L\theta_p)^2+\sigma_{\rm TB}^2$. Here, $\theta_p$ is the pointing error angle, and $\sigma_{\rm TB}^2$ is the variance due to atmospheric effects. The coupling efficiency $\eta(r)$ for a beam displaced by $r$ is described by \cite{vasylyev2012toward}:
\begin{equation}
\begin{aligned}
    &\eta(r) = \\
&\frac{2}{\pi W^2}\exp\left(-2\frac{r^2}{W^2}\right) \cdot \int_0^a \rho\exp\left(-2\frac{\rho^2}{W^2}\right)I_0\left(\frac{4r\rho}{W^2}\right) d\rho,
\end{aligned}
\end{equation}
where $W=W_{ST}(z=L)$ is the beam waist, $a$ is the receiver aperture radius, and $I_n(x)$ is the n-th order modified Bessel function. Additionally, we use a more efficient approximate analytical expression of $\eta(r)$, which is
\begin{equation}
    \eta(r) = \eta_0 \exp\left [-\left (\frac{ r}{R}\right )^\lambda\right ].
\end{equation}
In the expression, we have
\begin{equation}
\begin{aligned}
 &\eta_0 =1-\exp\left [-2\frac{a^2}{W^2}\right ],\\
&\lambda = \left [\ln\left (\frac{2T_0^2}{1-\exp[-4 \frac{a^2}{W^2}] {I}_0\left (4\frac{a^2}{W^2}\right )}\right )\right ]^{-1} \\
&\times 8\frac{a^2}{W^2} \frac{\exp\left [-4\frac{a^2}{W^2}\right ]{I}_1\left (4\frac{a^2}{W^2}\right )}{1-\exp[-4 \frac{a^2}{W^2}]{I}_0\left (4\frac{a^2}{W^2}\right )},\\
&R=a \left [\ln\left (\frac{2T_0^2}{1-\exp[-4 \frac{a^2}{W^2}]
{I}_0\left (4\frac{a^2}{W^2}\right )}\right )\right ]^{-\frac{1}{\lambda}}.
\end{aligned}
\end{equation}

\subsection{DM-CV KGP}

Additionally, excess noise in DM-CV type KGP is due to channel noise (scintillation, ambient light) and detection noise (device imperfections, electronic noise). Channel noise fluctuates based on environmental conditions and can be approximated by stray light power. Detection noise arises from inherent device flaws. Total excess noise is given by $\xi = \xi_{\rm ch} + \xi_{\rm det}/\eta$~\cite{Laudenbach2018Continuous}. Notably, heterodyne detection noise is twice that of homodyne detection~\cite{Li2024Discrete}.

\subsubsection{Key Rate Estimation for Heterodyne Detection}
The key rate estimation in this section refers to the numerical method presented in \cite{tan2021computing,coles2012unification}. The entire system consists of three parties: Alice, Bob and Eve. The environment Eve ensures that the combined system forms a global pure state. We divide Alice's system into two parts: a measurement result register $Z$ and a remaining quantum state $A$. For the secure key rate, which is mainly given by the conditional von Neumann entropy $H(Z|E)$, we have:
$$
H(Z|E) = H(ZE) - H(E)
$$
Since the global state $ZABE$ is pure, the von Neumann entropy of a subsystem is equal to the entropy of its complement. Thus, we can write
$$
H(Z|E) = H(ZE) - H(E) = H(AB) - H(ZAB)
$$
Based on \cite{coles2012unification}, the conditional entropy can be expressed as a quantum relative entropy
$$
H(Z|E) = D(\mathcal{G}(\rho_{AB})||\mathcal{Z}(\mathcal{G}(\rho_{AB})))
$$
where $D(\rho||\sigma) = \text{Tr}(\rho \log \rho) - \text{Tr}(\rho \log \sigma)$ is the quantum relative entropy. $\mathcal{G}$ represents the measurement operation and $\mathcal{Z}$ named the pinching channel. For heterodyne detection, the measurement $\mathcal{G}(\rho)=K\rho K^\dagger$ performed by Bob in the protocol can be described by Kraus operators
$$
K = \sum_{z=0}^3 \ket{z}_Z \otimes \mathbbm{1}_A \otimes (\sqrt{R_z})_B,
$$
where $\ket{z}_R$ represents the measurement outcome in a classical register, $\mathbbm{1}_A$ is the identity operator on Alice's system, and $(\sqrt{R_z})_B$ acts on Bob's system. The projection operator $R_z$ is given by:
$$
R_z = \frac{1}{\pi} \int_{\Delta_a}^\infty d\gamma \int_{z\pi/2+\Delta_p}^{(z+1)\pi/2-\Delta_p} \gamma\ket{\gamma e^{i\theta}}\bra{\gamma e^{i\theta}}d\theta.
$$
Here, $\ket{\gamma e^{i\theta}}$ denotes a coherent state. The parameters $\Delta_a$ and $\Delta_p$ define the integration range for amplitude and phase, respectively.
The corresponding pinching channel $\mathcal{Z}(\rho)$ can be represented using projection operators $Z_j = \ket{j}\bra{j}\otimes \mathbbm{1}_{AB}$ as:
$$
\mathcal{Z}(\rho) = \sum_{j=0}^{3} Z_j \rho Z_j.
$$
To determine the lower bound of the key rate, we need to find the minimum relative entropy over all possible states $\rho_{AB}$ that satisfy certain constraints. Considering error correction, the asymptotic key rate can be expressed as:
$$
R^\infty = \min_{\rho_{AB} \in S} D(\mathcal{G}(\rho_{AB})||\mathcal{Z}[\mathcal{G}(\rho_{AB})]) - p_{\rm pass}\delta_{EC}.
$$
The set $S$ of allowed states $\rho_{AB}$ must satisfy the following constraint conditions, derived from randomly sampled and publicly announced measurement data~\cite{liu2021homodyne,lin2019Asymptotic}
\begin{align}
    \text{Tr}[\rho_{AB}(\ket{x}\bra{x}_A\otimes \hat{q})] &= p_x \braket{\hat{q}}_x, \label{eq:constraint1}\\
    \text{Tr}[\rho_{AB}(\ket{x}\bra{x}_A\otimes \hat{p})] &= p_x \braket{\hat{p}}_x, \label{eq:constraint2}\\
    \text{Tr}[\rho_{AB}(\ket{x}\bra{x}_A\otimes \hat{n})] &= p_x \braket{\hat{n}}_x, \label{eq:constraint3}\\
    \text{Tr}[\rho_{AB}(\ket{x}\bra{x}_A\otimes \hat{d})] &= p_x \braket{\hat{d}}_x. \label{eq:constraint4}
\end{align}
Here, $p_x$ is the probability of Alice sending the corresponding quantum state $\ket{x}$, which is set to $p_x=0.25$. The expectation values are given by:
\begin{align}
    \braket{\hat{q}}_x &= \sqrt{2\eta}\text{Re}(\alpha_x), \label{eq:q_exp}\\
    \braket{\hat{p}}_x &= \sqrt{2\eta}\text{Im}(\alpha_x), \label{eq:p_exp}\\
    \braket{\hat{n}}_x &= \eta|\alpha_x|^2+\frac{\eta\xi}{2}, \label{eq:n_exp}\\
    \braket{\hat{d}}_x &= \eta[\alpha_x^2+(\alpha_x^*)^2].    \label{equation_qpnd}
\end{align}
In these equations, $\eta$ represents the detection efficiency, $\xi$ is the electronic noise, and $\alpha_x$ is the complex amplitude of the coherent state prepared by Alice. Operators $\hat{q}$ and $\hat{p}$ are the quadrature operators, $\hat{n}$ is the photon number operator, and $\hat{d}$ is related to the second moment of the quadratures.
Additionally, since Alice prepares the quantum states for Bob, Alice's reduced density matrix should have a fixed distribution:
$$
\text{Tr}_B[\rho_{AB}] = \sum_{i,j=0}^{3} \sqrt{p_i p_j} \braket{\phi_j|\phi_i}\ket{i}\bra{j}_A.
$$
Here, $\ket{\phi_i}$ are the coherent states prepared by Alice.
Finally, as a valid density matrix, $\rho_{AB}$ must satisfy the following conditions:
\begin{align}
        \text{Tr}[\rho_{AB}] &= 1, \label{eq:trace_condition}\\
        \rho_{AB} &\ge 0. \label{eq:positive_semidefinite}
\end{align}
The condition $\rho_{AB} \ge 0$ means that $\rho_{AB}$ must be a positive semi-definite operator.
The error rate needs to be considered. For given parameters, the probability of Bob obtaining measurement outcome $z=j$ given Alice sent state $x=k$ can be predicted by~\cite{lin2019Asymptotic}
\begin{align}
    &P(z=j|x=k) = \text{Tr}(R_j\rho_B^k) \\
    &= \int_{\Delta_a}^{\infty}d\gamma\int_{j\pi/2+\Delta_p}^{(j+1)\pi/2-\Delta_p}\frac{\exp\left(-\frac{|\gamma e^{i\theta}-\sqrt{\eta}\alpha_k|^2}{1+\eta\xi/2}\right)}{\pi\left(1+\eta\xi/2\right)}\gamma d\theta.
\end{align}
The parameters for the error correction term can then be calculated as~\cite{liu2021homodyne}
\begin{align}
p_{\rm pass} &= \sum_{x}p_x[P(0|x)+P(1|x)], \\
\delta_{EC} &= (1-\beta)H(Z) + \beta H(Z|X), \\
H(Z|X) &= \sum_{x}p_x h\left(\frac{P(0|x)}{P(0|x)+P(1|x)}\right).
\end{align}
In these equations, $\beta$ is the error correction efficiency, set to $0.95$. $H(Z)$ is the Shannon entropy of Bob's outcomes, and $I(X;Z)$ is the mutual information between Alice's inputs and Bob's outcomes. $h(x)$ is the binary entropy function, defined as $h(x) = -x \log_2 x - (1-x) \log_2 (1-x)$.

\subsubsection{Key Rate Estimation for Homodyne Detection}
Similarly, for homodyne detection, a similar key rate estimation approach is used. The main distinction is that homodyne detection can measure two quadratures, $q$ and $p$. In our protocol, these two measurement types occur with equal probability, so we average their contributions:
\begin{equation}
\begin{aligned}
        R^\infty &=  \frac{1}{2} \{\min_{\rho_{AB} \in S} \sum_{y\in \{q,p\}} D(\mathcal{G}_y(\rho_{AB})||\mathcal{Z}[\mathcal{G}_y(\rho_{AB})]) \\
        &-\sum_{y\in \{q,p\}}p^y_{\rm pass}\delta_{\rm EC}^y\}.
\end{aligned}
\end{equation}
For each quadrature measurement $y \in \{q,p\}$, the post-processing mapping $\mathcal{G}_y(\rho) = K_y \rho K_y^\dag$ is defined by the Kraus operators:
\begin{equation}
    K_y  = \sum_{b=0}^{1} \ket{b}_Z \otimes \mathbbm{1}_A \otimes (\sqrt{I_y^b})_B.
\end{equation}
In this expression, $\{I_y^b\}$ denotes a set of projective operators:
\begin{equation}
    I_y^0 = \int_{\Delta_c}^\infty dt\ket{t}\bra{t}, \quad I_y^1= \int_{-\infty}^{-\Delta_c} dt \ket{t} \bra{t}.
\end{equation}
Here, $\ket{t}$ represents an eigenstate of the quadrature operator $\hat{y}$, and $\Delta_c$ is a threshold value for the measurement outcome.
Similarly, the error rate can be calculated. The probability distributions for the quadrature outcomes $q$ and $p$ given Alice's input $\alpha_x$ are:
\begin{equation}
\begin{split}
    P(q|x) &= \frac{1}{\sqrt{\pi(\eta\xi+1)}}\exp\left(-\frac{(q-\sqrt{2\eta}\text{Re}(\alpha_x))^2}{\eta\xi+1}\right),\\
    P(p|x) &= \frac{1}{\sqrt{\pi(\eta\xi+1)}}\exp\left(-\frac{(p-\sqrt{2\eta}\text{Im}(\alpha_x))^2}{\eta\xi+1}\right).
\end{split}
\end{equation}
The probability distribution of Bob's bits for each quadrature $y \in \{q,p\}$ is expressed as:
\begin{equation}
\begin{aligned}
        &P_y(0|x) = \int_{\Delta_c}^{\infty}P(y|x)dy,\\
        &P_y(1|x) = \int_{-\infty}^{-\Delta_c}P(y|x)dy,\\
        &P_y(\perp|x) = \int_{-\Delta_c}^{\Delta_c}P(y|x)dy.
\end{aligned}
\end{equation}
Here, $P_y(0|x)$ is the probability of Bob obtaining outcome `0' for quadrature $y$ given Alice sent state $x$, $P_y(1|x)$ is the probability of obtaining `1', and $P_y(\perp|x)$ is the probability of a discarded result.

\subsection{BB84 KGP}
Following~\cite {lim2014concise}, the number of vacuum and single-photon events in the $X$-basis are estimated as:
\be\label{eqn2}
s_{X,0} \geq \tau_{0}\frac{\mu_\rd n_{X,\mu_\rdd}^--\mu_\rdd n_{X,\mu_\rd}^+}{\mu_\rd-\mu_\rdd},
\ee
\begin{equation} \label{eqn3}
s_{X,1} \geq \frac{\tau_{1}\mu_\rs\left[n_{X,\mu_\rd}^--n_{X,\mu_\rdd}^+-\frac{\mu_\rd^2-\mu_\rdd^2}{\mu_\rs^2}(n_{X,\mu_\rs}^+- \frac{s_{X,0}}{\tau_0})\right]}{\mu_\rs(\mu_\rd-\mu_\rdd)-\mu_\rd^2+\mu_\rdd^2}.
\end{equation}
where $\tau_{n}:=\sum_{k\in\cK}e^{-k}k^np_k/n!~(\cK = \{ \mu_1,\mu_2,\mu_3 \})$ is the probability of transmitting an $n$-photon state, and
\[
n_{X,k}^\pm:=\frac{e^{k}}{p_k}\left(n_{X,k}\pm\sqrt{ \frac{n_X}{2}\log\frac{21}{\esec}}\right),~\forall~k \in \cK.
\]
Here, $\mu_1$, $\mu_2$ and $\mu_3$ represent the signal and two decoy intensities in decoy-state BB84 type KGP. 

Equations analogous to these are used to calculate the number of vacuum events ($s_{Z,0}$) and single-photon events ($s_{Z,1}$) for $\cZ=\cup_{k\in\cK}\cZ_k$ using $Z$-basis statistics. The $X$-basis single-photon phase error rate is:
\be \label{eqn5}
\phi_{X,1}:=\frac{c_{X,1}}{s_{X,1}}  \leq \frac{v_{Z,1}}{s_{Z,1}} + \gamma^{U}\left( s_{Z,1},{s}_{X,1},\frac{v_{Z,1}}{s_{Z,1}},\esec \right), \ee
where
\[
v_{Z,1} \leq \tau_{1}\frac{m_{Z,\mu_\rd}^+-m_{Z,\mu_\rdd}^-}{\mu_\rd-\mu_\rdd},
\]
\[
m_{Z,k}^{\pm}:=\frac{e^{k}}{p_k}\left(m_{Z,k}\pm\sqrt{ \frac{m_Z}{2}\log\frac{21}{\esec}}\right),~\forall~k \in \cK.
\]

The total number of $X$-basis events is $n_X = \sum_{k \in \mathcal{K}} n_{X,k}$ and the number of errors is $m_X = \sum_{k \in \mathcal{K}} m_{X,k}$. 
Finally, the key rate is calculated by
    \begin{equation}
\begin{aligned}
    l_{\rm{key}} &= \underline{s}_{X,0} + \underline{s}_{X,1} \left[ 1 - H(\overline{\phi}_{X,1}^n) \right] \\
    &- \lambda_{EC}-\log_2(\frac{2}{\varepsilon_{\rm{cor}}}) - 6\log_{2}{\frac{22}{\epsilon_{PA}}}.
\end{aligned}
\end{equation}
where $\lambda_{EC}$ is the number of bits revealed during the error correction step, $\varepsilon_{\rm{cor}}$ is the failure probability of error correction, and $\epsilon_{PA}$ is the failure probability of privacy amplification.


%

\end{document}